\documentclass[times,preprint]{elsarticle}

\usepackage[english]{babel} 
\usepackage[T1]{fontenc}
\usepackage[utf8x]{inputenc}

\usepackage{jcomp}
\usepackage{framed,multirow}

\usepackage{amssymb}
\usepackage{latexsym}

\usepackage{url}
\usepackage{xcolor}
\definecolor{newcolor}{rgb}{.8,.349,.1}

\journal{Journal of Computational Physics}

\usepackage{amsmath}
\usepackage{algorithm}
\usepackage{algorithmic}

\usepackage[pdftex]{hyperref} 
\hypersetup{colorlinks=true}
\hypersetup{linkcolor=black}
\hypersetup{citecolor=black}

\usepackage{caption}
\usepackage{subcaption}

\usepackage{tikz}
\usepackage{pgfplots}
\pgfplotsset{compat=1.5.1}
\usetikzlibrary{calc}
\usetikzlibrary{arrows}
\usetikzlibrary{patterns}

\begin{document}
	
\verso{C. Rettinger \textit{et al.}}
	
\begin{frontmatter}
	
	\title{An efficient four-way coupled lattice Boltzmann - discrete element method for fully resolved simulations of particle-laden flows}%
	
	\author[1]{Christoph \snm{Rettinger}\corref{cor1}}
	\cortext[cor1]{Corresponding author}
	\ead{christoph.rettinger@fau.de}
	\author[1,2]{Ulrich \snm{Rüde}}
	
	\address[1]{Chair for System Simulation, Friedrich--Alexander--Universität Erlangen--Nürnberg, Cauerstraße 11, 91058 Erlangen, Germany}
	\address[2]{CERFACS, 42 Avenue Gaspard Coriolis, 31057 Toulouse Cedex 1, France}
	
	\received{...}
	\finalform{...}
	\accepted{...}
	\availableonline{...}
	\communicated{...}
	
	
\begin{abstract}
	A four-way coupling scheme for the direct numerical simulation of particle-laden flows is developed and analyzed.
	It employs a novel adaptive multi-relaxation time lattice Boltzmann method to simulate the fluid phase efficiently.
	The momentum exchange method is used to couple the fluid and the particulate phase.
	The particle interactions in normal and tangential direction are accounted for by a discrete element method using linear contact forces.
	All parameters of the scheme are studied and evaluated in detail and precise guidelines for their choice are developed.
	The development is based on several carefully selected calibration and validation tests of increasing physical complexity.
	It is found that a well-calibrated lubrication model is crucial to obtain the correct trajectories of a sphere colliding with a plane wall in a viscous fluid.
	For adequately resolving the collision dynamics it is found that the collision time must be stretched appropriately.
	The complete set of tests establishes a validation pipeline that can be universally applied to other fluid-particle coupling schemes	providing a systematic methodology that can guide future developments.
\end{abstract}

\end{frontmatter}


\section{Introduction}

With the increasing performance of today's computer hardware and supercomputers, simulations of particulate flows are becoming an increasingly popular tool for engineers to investigate and predict the rich dynamics of such systems. 
Depending on the usage of a macroscopic or microscopic description of the system~\cite{vanDerHoef2008,wachs2019}, the available simulation approaches differ substantially in terms of accuracy and computational cost.
Among them, the class of Direct Numerical Simulations (DNS) provides unique access to detailed force information of single grains or flow measurements inside small pores~\cite{wachs2019}.
This is achieved by fully resolving all flow features and particle shapes.
Such four-way coupled simulation approaches explicitly account for fluid-particle, particle-fluid, and particle-particle interactions.
As such, they promote a more in-depth insight into the complex interactions of the fluid and particle phase than laboratory experiments, which makes them a valuable tool to develop a better understanding of the underlying processes.

Recent application examples of this methodology can be found in the study of the effect of particles on turbulent flows~\cite{costa2018,costa2020}, dynamics of fluidization phenomena~\cite{esteghamatian2017,derksen2019}, or sediment erosion~\cite{kidanemariam2014interface,vowinckel2016}.
Most commonly, a discretization of the Navier-Stokes equations for the fluid flow is coupled via the immersed boundary method (IBM) to a soft-contact collision model, which accounts for particle interactions~ \cite{costa2018,costa2020,kidanemariam2014interface,vowinckel2016}.
We will refer to methods that directly discretize the Navier-Stokes equations, collectively as \textit{classical DNS approaches}.

Especially for dense particulate systems, where a large number of particle collision can be expected, a careful and accurate treatment of these particle interactions is necessary to obtain the correct system dynamics \cite{kempe2014relevance}.
Therefore, detailed studies have been conducted to calibrate and validate the classical DNS approaches~\cite{breugem2010,kempe2012,costa2015,biegert2017,jain2019}.
For that purpose, they commonly make use of experimental data featuring the collision dynamics of a single sphere with a wall~\cite{gondret2002}.
To match these experimental findings, different adaptions to the original coupling scheme have been proposed.
One example is the stretching of the collision event in time such that the contact time is much larger than predicted by contact theory to accurately resolve the collision event in time~\cite{breugem2010,kempe2012,costa2015}.
Other changes are temporal substepping for the particle simulation~\cite{biegert2017}, lubrication correction models that account for unresolved hydrodynamic interactions~\cite{breugem2010}, and disabling the hydrodynamic interaction force~\cite{kempe2012,costa2015,biegert2017} and adapting the IBM interpolation kernel~\cite{costa2015,biegert2017} during the collision.

As a promising alternative to the classical DNS approaches, the lattice Boltzmann method (LBM) can be employed to simulate the fluid flow~\cite{aidun2010,krueger2017}.
Besides an LBM adaptation of the immersed boundary method~\cite{feng2004}, other LBM-specific coupling approaches are available, like the momentum exchange method~\cite{ladd1994,aidun1998} or the partially saturated cells method~\cite{noble1998}.
Due to the LBM's benefits regarding parallelization and in combination with appropriate particle interaction models, such approaches are successfully applied to study complex and large-scale particulate systems~\cite{rettinger2017Riverbed,yang2017,seil2018,derksen2019,benseghier2020}.
In remarkable contrast to the aforementioned classical DNS approaches, however, we are not aware of numerical studies that rigorously assess the collision treatment of particles in viscous fluids using the LBM.
Setups without particle collision, like the case of a single settling sphere inside a fluid-filled box~\cite{tenCate2002}, are typically considered as validation scenarios~\cite{yang2017,seil2018}. 

To close this apparent gap and study whether algorithmic adaptations similar to the ones for the classical DNS approaches are required also for LBM coupling schemes, the present work presents an in-depth calibration and validation study of such a four-way coupled simulation approach.
We employ the momentum exchange method to establish the fluid-particle coupling whose benefits in terms of accuracy have already been shown previously~\cite{rettinger2017}.
We develop an adaptive multi-relaxation time LBM to reduce the spurious density fluctuations that are inherently present in this coupling scheme~\cite{peng2016}. 
The particle collisions are treated as soft-contacts with the discrete element method (DEM) due to its modularity and extensibility.
Those parts are carefully selected to construct a combined scheme that offers accurate predictions of various tests.
At the same time, we keep the number of parameters that all have to be calibrated, as low as possible.

Accordingly, the present work is an encompassing and detailed description of this approach and an exploration of its limitations.
Furthermore, we identify all parameters of the scheme and establish guidelines for their choice via calibration studies.
This requires that the corresponding test setups are chosen in such a way that the effect of a single parameter can be extracted to avoid ambiguity.
Consequently, we naturally establish a pipeline of well-documented \textit{acceptance} tests with increasing complexity in the underlying physics and the required numerical models.  
Most of these tests are not specific to LBM DNS approaches and can thus be readily applied to develop, calibrate, and validate other existing coupling schemes in a structured way.
This goes hand in hand with the observation and our experience that identifying appropriate tests and simulation setups is often time-consuming. 
Note that we restrict ourselves to three-dimensional setups to avoid artificial effects originating from two-dimensional simplifications that have to be counteracted by e.g. introducing a hydraulic radius~\cite{boutt2007}.   

All parts of the coupled algorithm are carefully selected to adhere to the parallelization and performance principles that have been developed and investigated in previous works~\cite{goetz2010,bartuschat2018,rettinger2017Riverbed}.
This is crucial to exploit the compute power of modern supercomputers fully and to carry out massively parallel simulations for large problem sizes efficiently.

\begin{figure}[t]
	\centering
	\begin{tikzpicture}
	
	\draw[fill=black!15!white] (-0.65,-0.9) rectangle (14.25,3.15);
	\draw[fill=black!10!white] (-0.45,-0.7) rectangle (10.5,2.95);
	\draw[fill=black!5!white] (-0.25,-0.5) rectangle (6,2.75);
	
	\draw[step=0.5,draw=black,thin,fill=cyan!30!white] (0,0) grid (2.5,2.5) rectangle (0,0);
	
	
	\draw[-latex,thick] (2.75,1.1) -- ++(2.5,0);
	\draw[-latex,thick] (5.25,1.4) -- ++(-2.5,0);
	
	\draw[draw=orange,very thick,fill=orange!20!white] (6.25,1.25) circle (0.75);
	
	\draw[-latex,thick] (7.25,1.1) -- ++(2.5,0);
	\draw[-latex,thick] (9.75,1.4) -- ++(-2.5,0);
	
	
	\draw[draw=orange,very thick,fill=orange!20!white] (10.75,1.25) circle (0.75);
	\draw[draw=orange,very thick,fill=orange!20!white] (12.25,1.25) circle (0.75);
	\fill[very thick,fill=orange!20!white] (13,0.5) rectangle (14,2);
	\draw[orange,very thick] (13,0.5) -- ++(0,1.5);
	\draw[-latex,thick] (11,1.1) -- ++(1,0);
	\draw[-latex,thick] (12,1.4) -- ++(-1,0);
	\draw[-latex,thick] (12.5,1.1) -- ++(1,0);
	\draw[-latex,thick] (13.5,1.4) -- ++(-1,0);
	
	\node[above] at (1.25,-0.5) {fluid flow};
	\node[above,align=center] at (4,-0.5) {fluid-particle\\interaction};
	\node[above,align=center] at (8.5,-0.5) {particle interaction\\via lubrication};
	\node[above,align=center] at (12.25,-0.5) {particle interaction\\via collisions};
	
	\node at (4,2.5) {Section \ref{sec:fluid_particle_interactions}};
	\node at (8.5,2.5) {Section \ref{sec:lubrication_interactions}};
	\node at (12.25,2.5) {Section \ref{sec:collisional_interactions}};
	
	\end{tikzpicture}
	\caption{Structure of the present work. Each section adds and focuses on a specific interaction of the final four-way coupling by presenting the numerical methods, and corresponding calibration and validation tests.}
	\label{fig:paperOverview}
\end{figure}
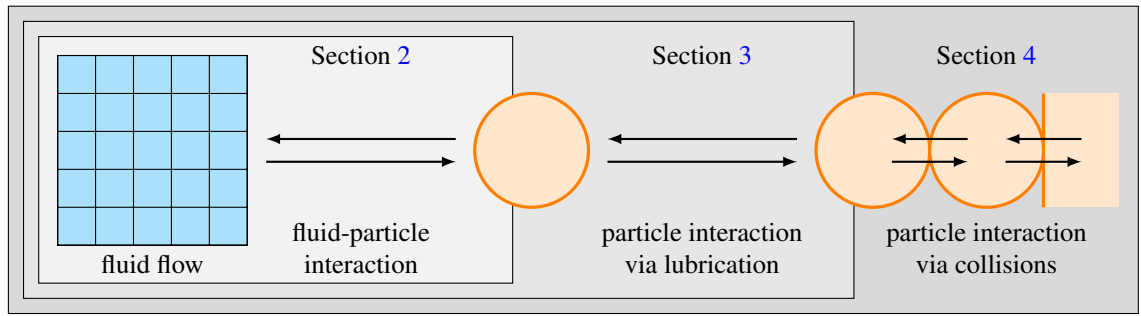

The ordering of the tests within this calibration and validation pipeline also motivates the structure of this work, as depicted in Fig.~\ref{fig:paperOverview}.
In the first part, Sec.~\ref{sec:fluid_particle_interactions}, we present our LBM for the fluid flow and the momentum exchange method together with tests involving particles that are either stationary or moving with a prescribed velocity.
In Sec.~\ref{sec:lubrication_interactions}, we present and evaluate our lubrication correction model that is crucial to represent hydrodynamic interactions between almost touching particles adequately.
We finalize our four-way coupling scheme in Sec.~\ref{sec:collisional_interactions} by accounting for particle collisions and permitting free particle motion.
This also contains extensive tests of normally and tangentially colliding spheres with a plane wall.
We summarize the main findings and conclude the work in Sec.~\ref{sec:conclusion}.
All particle specific definitions can be found in \ref{app:particleDefinitions}. 
All presented features and tests are implemented in the open-source \textsc{waLBerla} framework~\cite{godenschwager2013,bauer2020} and can be downloaded from the official repository\footnote{\url{www.walberla.net}}.

\section{Fluid-particle interaction}

\label{sec:fluid_particle_interactions}

\subsection{Fluid flow simulation with the lattice Boltzmann method}
\label{sec:lbm}

%

Having its origin in statistical mechanics, the lattice Boltzmann method describes the evolution of particle distribution functions (PDFs) on a uniform grid.
A general overview of the various aspects of the LBM can be found in the book of Kr\"uger \textit{et al.}~\cite{krueger2017}.
In the present article, we make use of the $D3Q19$ lattice model \cite{qian1992}, i.e., each cell of a three-dimensional grid contains $19$ PDFs, $f_q$, where each is associated with a specific discrete lattice velocity $\boldsymbol{c}_q$.
One time step of the lattice Boltzmann method is then split into two steps, the collision and the streaming step.
In the collision step, the PDFs are updated locally in each cell according to 
\begin{equation}
\tilde{f}_q(\boldsymbol{x},t) = f_q(\boldsymbol{x},t) + \mathcal{C}_q\left(f_1(\boldsymbol{x},t),...,f_{19}(\boldsymbol{x},t)\right),\label{eq:LBM_Collide}
\end{equation}
where $\mathcal{C}_q$ is a general collision operator that will be specified in Seq.~\ref{sec:mrt}. 
This is then followed by the streaming step, given as 
\begin{equation}
f_q( \boldsymbol{x} + \boldsymbol{c}_q\,\Delta t ,t+\Delta t) = \tilde{f}_q(\boldsymbol{x},t), \label{eq:LBM_Stream}
\end{equation}
which distributes the post-collision PDFs to neighboring cells.

The fluid density $\rho_f$ and velocity $\boldsymbol{u}_f$ are cell local quantities and obtained via moments of the PDFs:
\begin{equation}
\rho_f(\boldsymbol{x},t) = \sum_q f_q(\boldsymbol{x},t),\quad \boldsymbol{u}(\boldsymbol{x},t) = \frac{1}{\rho_0}\sum_q f_q(\boldsymbol{x},t) \boldsymbol{c}_q,
\end{equation}
where the mean density $\rho_0$ is introduced to reduce compressibility effects that are inherently present in the LBM~\cite{he1997}.
The pressure $p_f$ is directly connected to the density via
\begin{equation}
p_f = \rho_f c_s^2, \label{eq:pressure}
\end{equation}
where $c_s$ is the lattice speed of sound.

It can be shown that this numerical algorithm results in an approximation of the Navier-Stokes equations~\cite{krueger2017}.
In the context of LBM, it is common to express quantities in so-called lattice units which results in the cell size $\Delta x = 1$, the time step size $\Delta t = 1$, $\rho_0 = 1$, and $c_s = 1/\sqrt{3}$.
Those will be used in the remainder of this work.


\subsubsection{Multiple-relaxation-time collision model}
\label{sec:mrt}

A crucial part of LBM is the specific choice of the collision operator, which greatly influences the stability and accuracy of the fluid flow simulation.
The main principle is that during the collision step, the PDFs are relaxed towards an equilibrium state.
Most collision models can be generalized by considering the collision taking place in the moment space, in contrast to the velocity space containing the PDFs~\cite{dHumieres2002}.
There, the collision operator can be written as
\begin{equation}
\mathcal{C} = \mathbf{M}^{-1} \mathbf{S} \left(\boldsymbol{m}^{eq} - \mathbf{M}\boldsymbol{f}\right), \label{eq:MRT}
\end{equation}
with the moment transformation matrix $\mathbf{M}$, the relaxation rate matrix $\mathbf{S}$ and the vector $\boldsymbol{m}^{eq}$, containing the equilibrium moments.
This approach allows for relaxing different moments with different relaxation rates, and hence is called multiple-relaxation-time (MRT) model.

In particular, we here employ the MRT model of Ref.~\citenum{duenweg2007} for which the moment transformation matrix and the equilibrium moments are given in \ref{app:MRT}.
The diagonal matrix $\mathbf{S}$ contains the relaxation rates and can be written as
\begin{equation}
\mathbf{S} = \text{diag}\left( 0,\ 0,\ 0,\ 0, \  s_1, \  s_2, \  s_{3}, \  s_{3}, \  s_{3}, \  s_{4}, \  s_{4}, \  s_{4}, \  s_{4}, \  s_{4}, \  s_{5}, \  s_{5}, \  s_{6}, \  s_{6}, \  s_{6}\right), \label{eq:RelaxationRateMatrix}
\end{equation}
where the leading zeros originate from the conserved moments for mass and momentum, remaining unchanged during collision.
It can be shown that the fluid kinematic viscosity $\nu_f$ is determined by the relaxation rate of the shear moments, denoted as $s_\nu$, via
\begin{equation}
\nu_f = \tfrac{1}{3}\left(\tfrac{1}{s_{\nu}}-\tfrac{1}{2}\right) \label{eq:Viscosity}.
\end{equation}
Similarly, the bulk viscosity $\nu_b$ is given by the relaxation rate of the bulk moment, $s_b$, as 
\begin{equation}
\nu_b = \tfrac{2}{9}\left(\tfrac{1}{s_b}-\tfrac{1}{2}\right) \label{eq:BulkViscosity}.
\end{equation}
For the MRT model, as applied here, we have $s_4 = s_\nu$ and $s_1=s_b$~\cite{duenweg2007}. 
The remaining relaxation rates in $\mathbf{S}$ are not related to any macroscopic transport coefficients but can be used to increase numerical stability or accuracy without affecting the underlying physics.
The optimal choice, however, often depends on the specific problem at hand which then requires a time-consuming calibration procedure~\cite{krueger2017}.
It is thus desirable to decrease the number of free parameters in this model to the necessary minimum. 

\subsubsection{MRT specializations}

The general formulation of the collision operator in Eq.~\eqref{eq:MRT} allows us to describe common LBM variants conveniently as special cases.

The single-relaxation-time (SRT) or BGK model \cite{bhatnagar1954}, is recovered by setting all relaxation rates to the single relaxation rate from Eq.~\eqref{eq:Viscosity}, resulting in
\begin{equation}
\mathbf{S}^\text{SRT} = \text{diag}\left( 0,\ 0,\ 0,\ 0, \  s_\nu, \  s_\nu, \  s_\nu, \  s_\nu, \  s_\nu, \  s_\nu, \  s_\nu, \  s_\nu, \  s_\nu, \  s_\nu, \  s_\nu, \  s_\nu, \  s_\nu, \  s_\nu, \  s_\nu\right). \label{eq:SRTmatrix}
\end{equation}
Though conceptually simple and thus the most popular collision model, the SRT model has well-known drawbacks regarding stability and accuracy~\cite{krueger2017}.
In particular, it features an undesired dependency of the boundary location on its relaxation rate and thus on the fluid viscosity \cite{pan2006}.
This issue will be briefly revisited in Sec.~\ref{sec:StokesDrag}

The two-relaxation-time (TRT) model, developed by Ginzburg \textit{et al.}~\cite{ginzburg2008}, aims to overcome this drawback by introducing a second relaxation rate $s_\nu^-$.
This relaxation rate is used in Eq.~\eqref{eq:SRTmatrix} for all odd-order moments instead of $s_\nu$ and leads to
\begin{equation}
\mathbf{S}^\text{TRT} = \text{diag}\left( 0,\ 0,\ 0,\ 0, \  s_\nu, \  s_\nu, \  s_\nu^-, \  s_\nu^-, \  s_\nu^-, \  s_\nu, \  s_\nu, \  s_\nu, \  s_\nu, \  s_\nu, \  s_\nu, \  s_\nu, \  s_\nu^-, \  s_\nu^-, \  s_\nu^-\right). \label{eq:TRTmatrix}
\end{equation}
To obtain viscosity-independent boundary locations, the two relaxation rates have to satisfy the relation
\begin{equation}
\Lambda = \left(\frac{1}{s_\nu} - \frac{1}{2}\right)\left(\frac{1}{s_\nu^-} - \frac{1}{2}\right), \label{eq:MagicParameter}
\end{equation}
where $\Lambda$ is the so-called "magic" number, a constant, for which we use $\Lambda=3/16$~\cite{khirevich2015}. 

\subsubsection{TRT+B collision model}

In this work, we propose an extension to the TRT model, called TRT+B model, that allows us to control the bulk viscosity, Eq.~\eqref{eq:BulkViscosity}, independently while retaining the advantages of the TRT model.
This is achieved by explicitly introducing $s_b$ as the relaxation rate of the kinetic energy and kinetic energy square moments \cite{khirevich2015} and yields
\begin{equation}
\mathbf{S}^\text{TRT+B} = \text{diag}\left( 0,\ 0,\ 0,\ 0, \  s_b, \  s_b, \  s_\nu^-, \  s_\nu^-, \  s_\nu^-, \  s_\nu, \  s_\nu, \  s_\nu, \  s_\nu, \  s_\nu, \  s_\nu, \  s_\nu, \  s_\nu^-, \  s_\nu^-, \  s_\nu^-\right). \label{eq:TRTbulkmatrix}
\end{equation}
Even though it would suffice to only set $s_1=s_b$ to control the bulk viscosity, we follow the suggestion of Ref.~\citenum{khirevich2015} to avoid largely different values for $s_1$ and $s_2$ for stability reasons.
Instead of choosing $s_b$ completely independent of the other relaxation rates, we use the relation of Ref.~\citenum{khirevich2015}:
\begin{equation}
\Lambda_b = \frac{\left(\tfrac{1}{s_b}-\tfrac{1}{2}\right)}{\left(\tfrac{1}{s_\nu}-\tfrac{1}{2}\right)}. \label{eq:lambda_b}
\end{equation}
For $\Lambda_b=1$, the TRT+B model reduces to the standard TRT model, while $\Lambda_b > 1$ leads to an increase in the bulk viscosity. 
This relation ensures viscosity-independent boundary locations, as will be seen in Sec.~\ref{sec:StokesDrag}.

\subsection{Fluid-particle coupling with the momentum-exchange method}
\label{sec:coupling}

The two-way coupling of the fluid flow simulation with particles is here established via the LBM-specific momentum exchange method, originating from Ladd~\cite{ladd1994} and extended by Aidun \textit{et al.}~\cite{aidun1998}.
It applies boundary conditions for the fluid along the particle's surface and computes a hydrodynamic interaction force $\boldsymbol{F}_{p,i}^{fp}$ and torque $\boldsymbol{T}_{p,i}^{fp}$ that act on particle $i$.

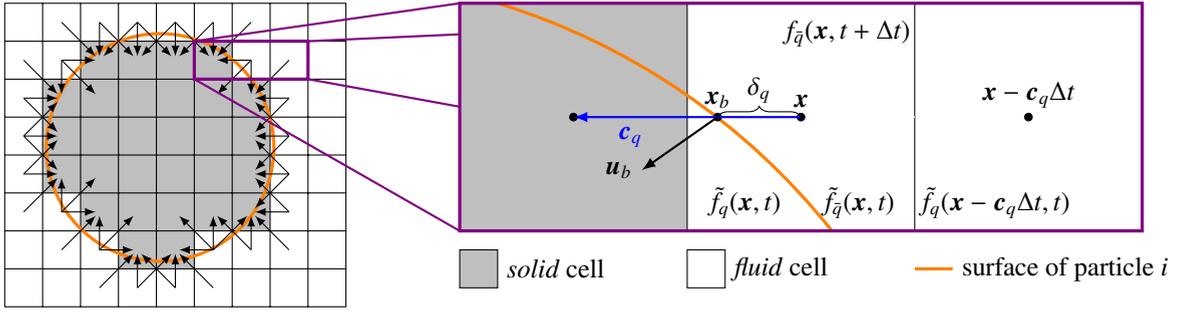
\begin{figure}[t]
	\centering
	\begin{tikzpicture}
	\coordinate[] (x1) at (2.05,2.1);
	
	\fill[lightgray] (1,1) rectangle ++(2,2.5);
	\fill[lightgray] (0.5,1.5) rectangle ++(0.5,1.5);
	\fill[lightgray] (1.5,0.5) rectangle ++(1,0.5);
	\fill[lightgray] (3,1) rectangle ++(0.5,2);
	
	\draw[step=0.5,black,thin] (0,0) grid (4.5,4);
	
	\draw[orange,very thick] (x1) circle (1.5);
	
	\def\arrowLength{0.45}
	\foreach \point in {(1.25,0.75),(0.75,1.25),(0.25,1.75),(0.25,2.25),(0.25,2.75),(0.75,3.25)}
	{
		\draw[->,-latex] \point -- ++(\arrowLength,0.0);
	}
	\foreach \point in {(1.25,0.75),(0.75,1.25),(1.75,0.25),(2.25,0.25),(2.75,0.75),(3.25,0.75)}
	{
		\draw[->,-latex] \point -- ++(0,\arrowLength);
	}
	\foreach \point in {(2.75,0.75),(3.75,1.25),(3.75,1.75),(3.75,2.25),(3.75,2.75),(3.25,3.25)}
	{
		\draw[->,-latex] \point -- ++(-\arrowLength,0);
	}
	\foreach \point in {(0.75,3.25),(1.25,3.75),(1.75,3.75),(2.25,3.75),(2.75,3.75),(3.25,3.25)}
	{
		\draw[->,-latex] \point -- ++(0,-\arrowLength);
	}
	\foreach \point in {(2.75,0.75),(1.75,0.25),(1.25,0.75),(1.25,0.25),(0.75,1.25),(0.75,0.75),(0.25,1.25),(0.25,1.75),(0.25,2.25)}
	{
		\draw[->,-latex] \point -- ++(\arrowLength,\arrowLength);
	}
	\foreach \point in {(3.25,0.75),(2.75,0.75),(2.75,0.25),(2.25,0.25),(3.75,1.25),(3.75,1.75),(3.75,2.25),(3.75,0.75)}
	{
		\draw[->,-latex] \point -- ++(-\arrowLength,\arrowLength);
	}
	\foreach \point in {(1.75,3.75),(2.25,3.75),(2.75,3.75),(3.25,3.75),(3.25,3.25),(3.75,3.25),(3.75,2.75),(3.75,2.25),(3.75,1.75)}
	{
		\draw[->,-latex] \point -- ++(-\arrowLength,-\arrowLength);
	}
	\foreach \point in {(0.25,2.25),(0.25,2.75),(0.25,3.25),(0.75,3.25),(0.75,3.75),(1.25,3.75),(1.75,3.75),(2.25,3.75)}
	{
		\draw[->,-latex] \point -- ++(\arrowLength,-\arrowLength);
	}

	\draw[violet,very thick] (2.5,3) rectangle ++(1.5,0.5);
	\draw[violet,thick] (2.5,3) -- (6,1);
	\draw[violet,thick] (2.5,3.5) -- (6,4);
	\draw[violet,thick] (4,3) -- (15,1);
	\draw[violet,thick] (4,3.5) -- (15,4);

	\fill[white] (6,1) rectangle ++(9,3);
	\fill[lightgray] (6,1) rectangle ++(3,3);
	\draw[step=3,black,thin,yshift=1cm] (6,0) grid ++(9,3);
	\draw[orange,very thick] (6.5,4) arc (73:38.5:9);

	\coordinate[] (x) at (10.5,2.5);
	\coordinate[] (xMcq) at (13.5,2.5);
	\coordinate[] (xb) at (9.4,2.5);
	\coordinate[] (xs) at (7.5,2.5);
	\coordinate[] (ub) at (-1,-0.7);
	
	\draw[white] (x) -- ($(x)!0.5!(xs)$) node[pos=0.5,anchor=north,yshift=-0.8cm] {\color{black}$\tilde{f}_q(\boldsymbol{x},t)$};
	\draw[white] (x) -- ($(x)!0.5!(xMcq)$) node[pos=0.5,anchor=north,yshift=-0.8cm] {\color{black}$\tilde{f}_{\bar{q}}(\boldsymbol{x},t)$};
	\draw[white] (xMcq) -- ($(xMcq)!0.3!(x)$) node[pos=0.5,anchor=north,yshift=-0.8cm] {\color{black} $\tilde{f}_q(\boldsymbol{x}-\boldsymbol{c}_q\Delta t,t)$};
	\draw[white] (x) -- ($(x)!0.5!(xMcq)$) node[pos=0.4,anchor=south,yshift=0.8cm] {\color{black}$f_{\bar{q}}(\boldsymbol{x},t+\Delta t)$};
	
	\draw[fill=black] (x) circle (0.05) node[anchor=south] {$\boldsymbol{x}$};
	\draw[fill=black] (xMcq) circle (0.05) node[anchor=south] {$\boldsymbol{x}-\boldsymbol{c}_q \Delta t$};
	\draw[fill=black] (xb) circle (0.05) node[anchor=south] {$\boldsymbol{x}_b$};
	\draw[fill=black] (xs) circle (0.05);
	
	\draw[|->,-latex,blue,thick] (x) -- (xs) node[pos=0.75,anchor=north] {$\boldsymbol{c}_q$};
	\draw [decorate,decoration={brace,amplitude=4pt}] (xb) -- (x) node [pos=0.5,anchor=south,yshift=2pt] {$\delta_q$};
	
	\draw[->,-latex,thick] (xb) -- ($(xb)+(ub)$) node[pos=1,anchor=east] {$\boldsymbol{u}_b$};
	
	\draw[violet,very thick] (6,1) rectangle ++(9,3); 
	
	\draw[fill=lightgray,thin] (6,0.25) rectangle ++(0.5,0.5);
	\node[right] at (6.5,0.5) {\textit{solid} cell};
	\draw[thin] (9,0.25) rectangle ++(0.5,0.5);
	\node[right] at (9.5,0.5) {\textit{fluid} cell};
	\draw[orange,very thick] (12,0.5) -- ++(0.5,0.0);
	\node[right] at (12.5,0.5) {surface of particle $i$};
	
	\end{tikzpicture}
	\caption{Sketch of the explicit particle mapping for the momentum-exchange method and all fluid-solid directions, shown as black arrows. The magnified area visualizes the CLI boundary condition, Eq.~\eqref{eq:MEM_CLI}. The stated PDF values $f_q$ are just given for illustration purposes and their position does not reflect the actual physical position.}
	\label{fig:MappingAndCLIsketch}
\end{figure}

A key aspect of this approach is the explicit mapping of the particles into the fluid domain.
Consequently, all cells with a cell center that is contained inside a particle are considered \text{solid} cells, carrying no fluid information, as opposed to \textit{fluid} cells that make up the domain where the fluid flow is solved for via the LBM~\cite{aidun1998}.
A sketch of this can be seen in the left part of Fig.~\ref{fig:MappingAndCLIsketch}.

The interaction with the fluid is then established by no-slip boundary conditions along the mapped surfaces of the particles.
Here, we use a higher order boundary condition called central linear interpolation (CLI) scheme~\cite{ginzburg2008}.
It uses subgrid information to obtain a better representation of the actual smooth particle surface which improves the accuracy of the coupling~\cite{rettinger2017}. 
This boundary condition is given by
\begin{equation}
f_{\bar{q}}(\boldsymbol{x},t+\Delta t) =  \tilde{f}_q(\boldsymbol{x},t) + \frac{1-2\delta_q}{1+2\delta_q} \left(\tilde{f}_q(\boldsymbol{x}-\boldsymbol{c}_q\Delta t, t) - \tilde{f}_{\bar{q}}(\boldsymbol{x},t)\right)-\frac{4}{1+2\delta_q}\frac{w_q\rho_0}{c_s^2}\boldsymbol{u}_b \cdot \boldsymbol{c}_q\ , \label{eq:MEM_CLI}
\end{equation}
where $\bar{q}$ denotes the opposite lattice direction of $q$, such that $-\boldsymbol{c}_q = \boldsymbol{c}_{\bar{q}}$, $w_q$ are the lattice weights~\cite{qian1992} and $\boldsymbol{u}_b = \boldsymbol{U}_i(\boldsymbol{x}_b)$ is the boundary velocity according to Eq.~\eqref{eq:particleVelocityAtPos}. 
The variable $\delta_q$ denotes the normalized distance of the cell center to the exact surface position such that $\boldsymbol{x}_b = \boldsymbol{x} + \delta_q\boldsymbol{c}_q\Delta t$.
An illustration of this boundary treatment and the accessed data is given in the right part of Fig.~\ref{fig:MappingAndCLIsketch}.
If not enough fluid information is available, i.e., the cell $\boldsymbol{x}-\boldsymbol{c}_q \Delta t$ is solid, we employ the bounce back condition that is obtained by setting $\delta_q=1/2$ in Eq.~\eqref{eq:MEM_CLI} and does not require neighbor information.

The resolved hydrodynamic force and torque acting on the particle are then evaluated with the momentum-exchange method~\cite{ladd1994}.
The local interaction force at a boundary location $\boldsymbol{x}_b$ and for a fluid-solid link $q$, i.e. a direction that has been treated by the boundary condition Eq.~\eqref{eq:MEM_CLI}, is given as~\cite{wen2014}
\begin{equation}
\boldsymbol{F}_q(\boldsymbol{x}_b,t) = \left(\boldsymbol{c}_q-\boldsymbol{u}_b\right) \tilde{f}_q(\boldsymbol{x},t) - \left(\boldsymbol{c}_{\bar{q}}-\boldsymbol{u}_b\right) f_{\bar{q}}(\boldsymbol{x},t+\Delta t). \label{eq:MEM_Fq}
\end{equation}
By summing up all fluid-solid directions $q$ and combining all contributions that originate from boundary locations of a specific particle $i$, i.e. all black arrows shown in Fig.~\ref{fig:MappingAndCLIsketch}, the currently acting hydrodynamic force and torque is obtained via~\cite{ladd1994}
\begin{align}
\boldsymbol{F}_{p,i}^{fp}(t) & = \sum_{\boldsymbol{x}_b\text{ of }i} \sum_q \boldsymbol{F}_q(\boldsymbol{x}_b,t)\ , \label{eq:interaction_force}\\
\boldsymbol{T}_{p,i}^{fp}(t) & = \sum_{\boldsymbol{x}_b\text{ of }i} \sum_q (\boldsymbol{x}_b - \boldsymbol{x}_{p,i}) \times \boldsymbol{F}_q(\boldsymbol{x}_b,t)\ . \label{eq:interaction_torque}
\end{align}
As noted in Ref.~\citenum{ernst2013}, some of these fluid-solid links might be missing when two surfaces are close to each other. Consequently, the momentum-balance of the fluid is incomplete which would result in an artificial attractive force between the two particles.
Here, we account for these missing force contributions by setting $\boldsymbol{F}_q(\boldsymbol{x}_b,t) = 2w_q\boldsymbol{c}_q$ in those cases which corresponds to the pressure force in stationary conditions.

Due to the explicit mapping of the particle into the domain, cells will eventually transition from solid to fluid.
In this case, valid fluid information has to be restored in the transition cell $\boldsymbol{x}_t$ before the simulation can continue, i.e., all PDF values have to be re-initialized in that cell.
Here, we employ the ideas presented in Refs.~\citenum{krithivasan2014} and \citenum{dorschner2015} that include local pressure tensor information to increase the accuracy, resulting in
\begin{equation}
f_q(\boldsymbol{x}_t) = w_q \left(\bar{\rho} + \frac{\bar{\rho} u_{t,\alpha} c_{q\alpha}}{c_s^2} + \frac{1}{2c_s^4}\left(\bar{\rho} u_{t,\alpha} u_{t,\beta} - \frac{\bar{\rho} c_s^2}{s_\nu}\left(\frac{\partial u_{t,\alpha} }{\partial x_\beta} + \frac{\partial u_{t,\beta}}{\partial x_\alpha }\right)\right)(c_{q\alpha}c_{q\beta} - c_s^2\delta_{\alpha\beta})\right), \label{eq:PDF_reconstruction}
\end{equation}
where $\delta_{\alpha\beta}$ denotes the Kronecker delta.
We use a spatially averaged density of all neighboring fluid cells $\bar{\rho}$, and the formerly present particle's velocity evaluated at $\boldsymbol{x}_t$, i.e. $\boldsymbol{u}_t=\boldsymbol{U}_i(\boldsymbol{x}_t)$.
The velocity derivatives are approximated by second-order finite differences which are replaced by first-order backward or forward finite differences if not enough neighboring fluid cells are available.

\subsection{Effect of bulk viscosity on the flow field near a moving particle}
\label{sec:BulkViscEffect}

\subsubsection{Background}

It is a well-known drawback of the momentum-exchange method that the hydrodynamic force that is acting on a moving particle and evaluated via Eqs.~\eqref{eq:interaction_force} and \eqref{eq:interaction_torque} exhibits oscillations~\cite{lallemand2003,peng2016}.
The main sources are the applied reconstruction algorithm in the wake of the particle, the varying number of cells that are used to represent the particle on the grid but also the general bounce-back-like formulation of LBM boundary conditions.
The latter one can be mitigated by averaging the force over two succeeding time-steps~\cite{ladd1994}, as will be explained in more detail in Sec.~\ref{sec:complete_algorithm}.
A careful choice of the reconstruction algorithm can also decrease these force oscillations~\cite{peng2016}, whereas the varying number of cells will always lead to visible grid effects on the force.

\begin{figure}[t]
	\centering
	\begin{subfigure}{0.22\textwidth}
		\centering
		\includegraphics[height=7cm]{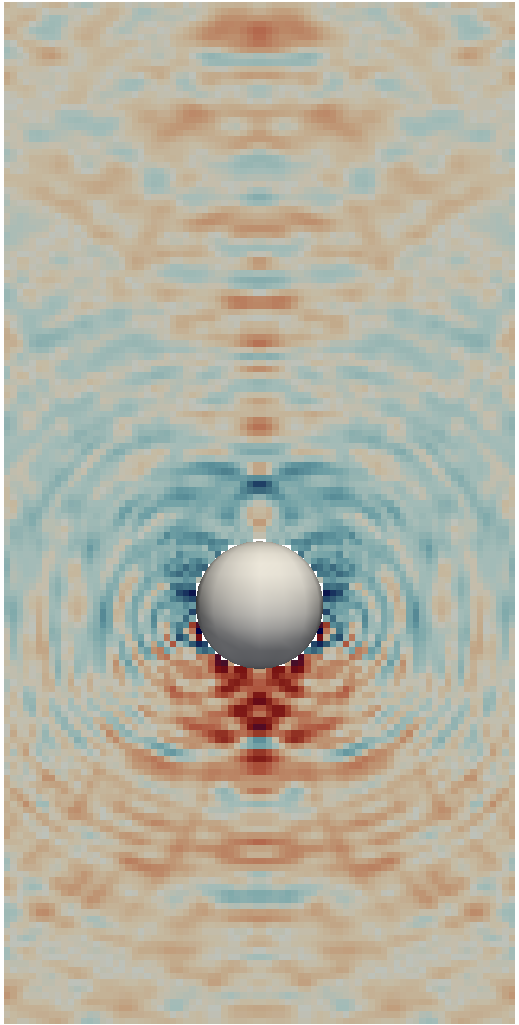}
		\caption{TRT ($\Lambda_b=1$).}
	\end{subfigure}
	~
	\begin{subfigure}{0.22\textwidth}
		\centering
		\includegraphics[height=7cm]{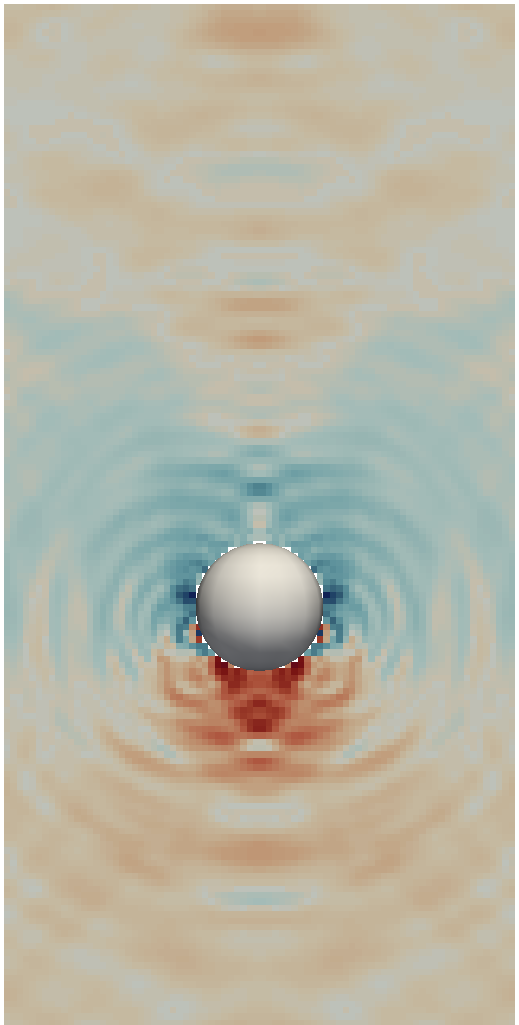}
		\caption{TRT+B ($\Lambda_b=10$).}
	\end{subfigure}
	~
	\begin{subfigure}{0.22\textwidth}
		\centering
		\includegraphics[height=7cm]{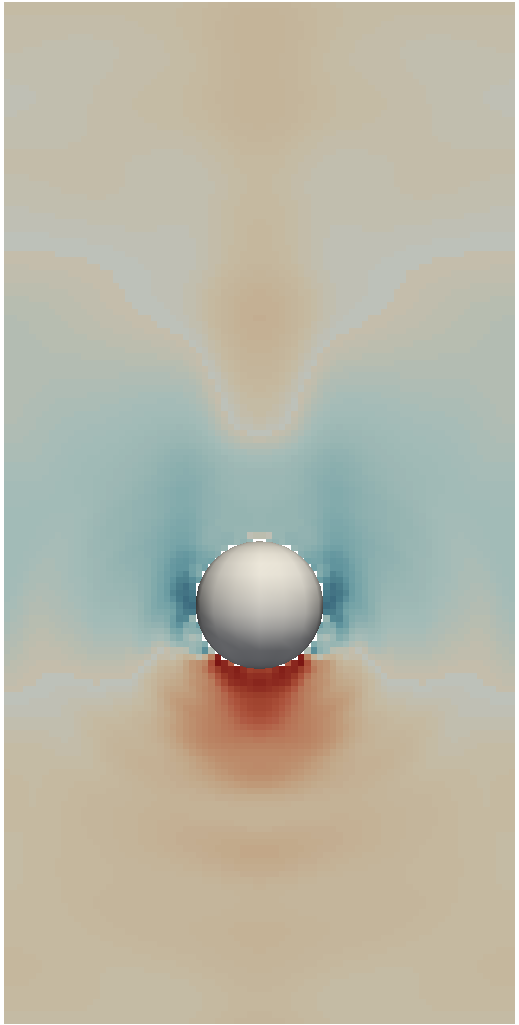}
		\caption{TRT+B ($\Lambda_b=100$).}
	\end{subfigure}
	~
	\begin{subfigure}{0.25\textwidth}
		\centering
		\includegraphics[height=7cm]{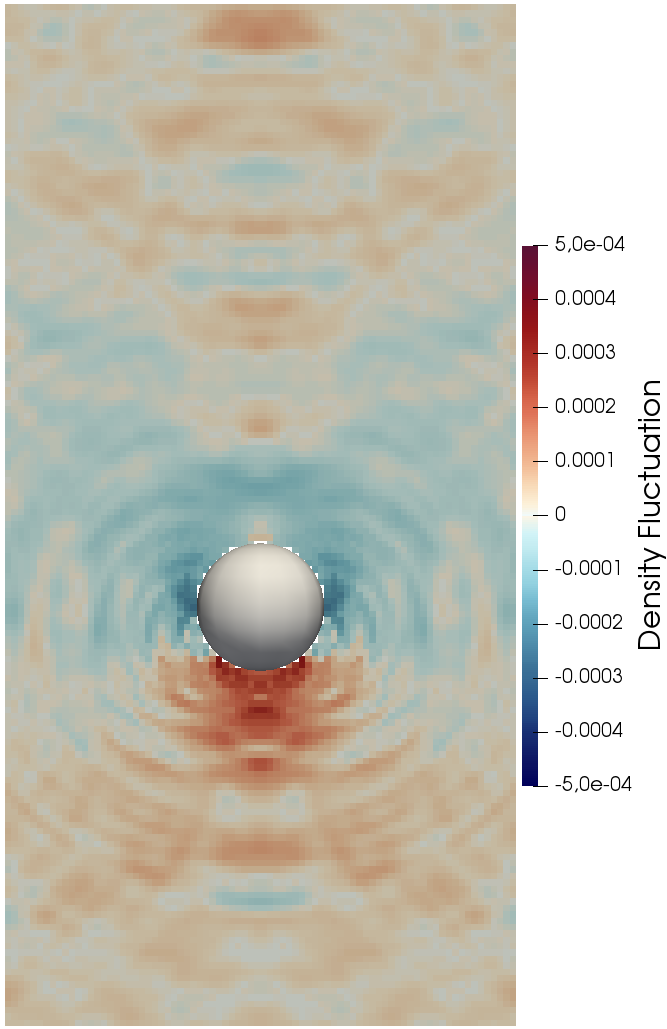}
		\caption{TRT+aB ($\Lambda_b=100$).}
	\end{subfigure}
	\caption{Effect of changing the bulk viscosity via the parameter $\Lambda_b$ on the density fluctuations originating from a spherical particle settling from top to bottom with a prescribed velocity.}
	\label{fig:effect_bulkvisc_vtk}
\end{figure}

It is less known, however, that also the flow field can exhibit visible disturbances originating from the moving particle.
This can be seen in Fig.~\ref{fig:effect_bulkvisc_vtk}(a) for a settling sphere, where the density fluctuations that clearly originate from the moving spherical particle and travel through the whole domain are visualized.
It can also be understood that such fluctuations will induce oscillations in the interaction force on the particle. 
In Refs.~\citenum{peng2016} and \citenum{tao2016}, these fluctuations have also been noted and have again been attributed to the reconstruction algorithm.
This is certainly one source but does not explain why these oscillations are most prominent in front of (in the figure: below) the particle and not in the wake, where one would expect the effect of the reconstruction.
Instead, we suspect that the update of the particle mapping, that will eventually but instantly convert fluid to solid cells, leads to a disturbance of the flow from the leading to the rear part of the particle and causes these observable fluctuations.
Since this mapping is an essential part of the momentum-exchange method, it cannot be changed easily and other ways to damp the oscillations have to be developed and evaluated.

\subsubsection{Adaptive TRT+B collision model}

\label{sec:trt_ab_model}

\begin{figure}[t]
	\centering
	\begin{tikzpicture}
	\coordinate[] (x1) at (3.3,2.4);
	\coordinate[] (x2) at (6.7,2.9);
	
	\fill[teal] (2,0.8) rectangle ++(2.4,3.2);
	\fill[teal] (1.6,1.6) rectangle ++(0.4,1.6);
	\fill[teal] (4.4,1.2) rectangle ++(0.4,2.4);
	\fill[teal] (4.8,1.6) rectangle ++(0.4,2);
	\fill[teal] (5.2,1.6) rectangle ++(0.4,2.4);
	\fill[teal] (5.6,1.2) rectangle ++(2,3.2);
	\fill[teal] (7.6,1.6) rectangle ++(0.4,2.8);
	\fill[teal] (8,2) rectangle ++(0.4,2);
	\fill[teal] (6,4.4) rectangle ++(1.2,0.4);
	
	\fill[lightgray] (2.4,1.6) rectangle ++(1.6,1.6);
	\fill[lightgray] (4,2) rectangle ++(0.4,0.8);
	\fill[lightgray] (6,2) rectangle ++(1.6,1.6);
	\fill[lightgray] (5.6,2.4) rectangle ++(0.4,0.8);
	\fill[lightgray] (6.4,3.6) rectangle ++(0.8,0.4);
	
	\draw[step=0.4,black,thin] (0,0.4) grid (10,5.2);
	
	\draw[orange,very thick] (x1) circle (1);
	\draw[orange,very thick] (x2) circle (1);
	\draw[black,thick,dashed] (x1) circle (1.8);
	\draw[black,thick,dashed] (x2) circle (1.8);
	
	\draw[<->,thick] (7.7,2.95) -- ++(0.8,0) node[anchor=north,pos=0.5,white]{$\boldsymbol{\delta_{aB}}$};
	
	\draw[thin] (10.8,4) rectangle ++(0.4,0.4);
	\node[right] at (11.3,4.2) {\textit{fluid} cell with $\Lambda_b=1$};
	\draw[fill=teal,thin] (10.8,3.2) rectangle ++(0.4,0.4);
	\node[right] at (11.3,3.4) {\textit{fluid} cell with $\Lambda_b>1$};
	\draw[fill=lightgray,thin] (10.8,2.4) rectangle ++(0.4,0.4);
	\node[right] at (11.3,2.6) {\textit{solid} cell};
	\draw[orange,very thick] (10.8,1.8) -- ++(0.4,0.0);
	\node[right] at (11.3,1.8) {particle surface};
	\draw[black, thick, dashed] (10.8,1) -- ++(0.4,0.0);
	\node[right] at (11.3,1) {bulk viscosity adaption shell};
	
	\end{tikzpicture}
	\caption{Sketch of the TRT+aB LBM collision model which increases the bulk viscosity via the parameter $\Lambda_b$ in all fluid cells that have their center inside a spherical shell of thickness $\delta_{aB}$ around each obstacle.}
	\label{fig:TRTaBsketch}
\end{figure}
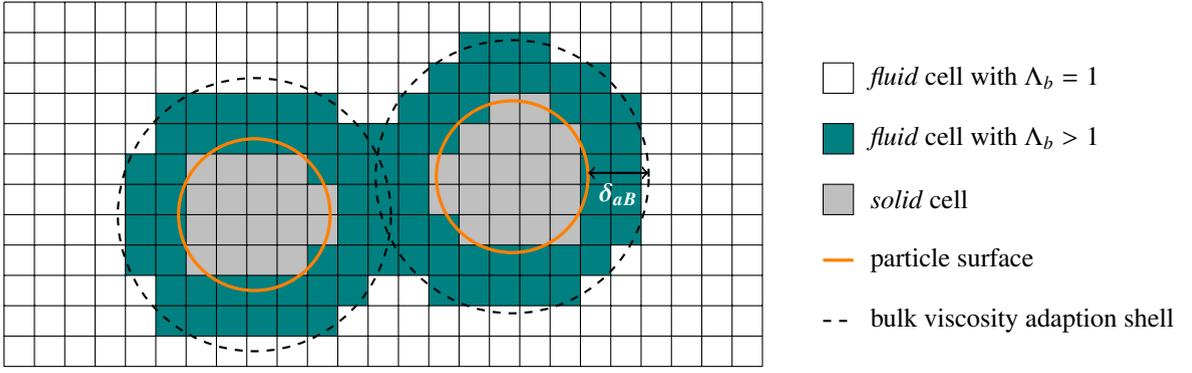

These spurious density oscillations can also be regarded as undesired compressibility effects of LBM that are not present in the incompressible flows under consideration in this work and which we must therefore strive to reduce.
This is our main motivation behind the introduction of the TRT+B collision model with the parameter $\Lambda_b$, Eq.~\eqref{eq:lambda_b}.
The thereby provided explicit control over the bulk viscosity of the fluid renders it a promising candidate to damp these density oscillations.

Since the fluctuations originate from the moving particle, we also propose and investigate a variant of the TRT+B model that only features an increased bulk viscosity in a spherical shell around particles, i.e., the source of the fluctuations, while at the same time maintaining $\Lambda_b=1$ elsewhere.
An illustration of this adaptive TRT+B model, which will be denoted as TRT+aB model, can be seen in Fig.~\ref{fig:TRTaBsketch}.
The thickness of the shell in which the bulk viscosity is adapted is chosen as $\delta_{aB} = 2\Delta x$.
This ensures that there is always at least one layer of cells around the particle with increased bulk viscosity.
At the same time, this keeps the particle information as local as possible, which is an essential feature for algorithms to be used in massively parallel environments like supercomputers.

\subsubsection{Description}

To investigate and evaluate the effect of the bulk viscosity on the density fluctuations, we construct the following settling test.
As such, it can be regarded as an LBM-specific test case as the bulk viscosity does not appear in the incompressible Navier-Stokes equations.
Consequently, such density, or pressure as stated in Eq.~\eqref{eq:pressure}, oscillations have not been reported for classical DNS approaches.

The test setup is chosen such that the flow and particle properties are very similar to the one that will be used later on in the test case of Sec.~\ref{sec:sphereWallCollision}.
It features a domain of size  $[4\times4\times8]D_p$, bounded by stationary walls in $x$- and $y$-direction and periodic in settling direction.
A sphere of diameter $D_p=20$ is initialized in the center of the domain, i.e. $\boldsymbol{x}_p(0)=D_p(2,2,4)^\top$.
To mimic an accelerating sphere, its $z$-velocity is prescribed in each time step according to
\begin{equation}
u_p(t) = - u_s \left(1 - \exp(-c_{acc}\,t/t_{\textit{St}})\right), \text{ with } t_{\textit{St}} = \tfrac{\rho_p}{\rho_f} \tfrac{D_p^2}{18\nu_f}  \label{eq:artificial_acceleration}
\end{equation}
and with an acceleration constant $c_{acc}$.
Here, we use $c_{acc}=10$ and $\rho_p/\rho_f=8.34$.
By prescribing the velocity, we can exclude other possible sources for oscillations, e.g., the feedback mechanism of having slightly oscillating hydrodynamic forces that are then transferred to the motion of the particle.
The flow conditions are fully specified by the Reynolds number $Re = u_sD_p/\nu_f = 164$, with $u_s = 0.02$ in lattice units.
To monitor the density fluctuations, a virtual probe is installed at location $\boldsymbol{x}_p(t) + (D_p,0,0)^\top$, i.e., it is moving with the particle and uses trilinear interpolation to retrieve the density value at the current probe location.
The simulation is run for $T = 24 t_{ref}$ with $t_{ref} = D_p/u_s$, resulting in slightly more than two passes through the whole domain.

\subsubsection{Results and Discussion}

\begin{figure}[t]
	\centering
	\input{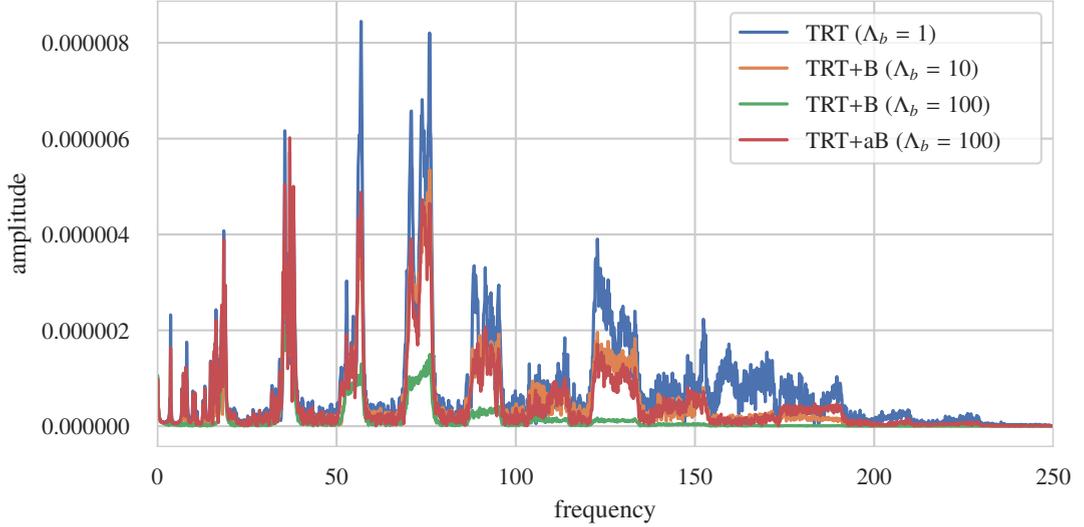}
	\caption{Spectrum of the density fluctuations at the probe location for the cases shown in Fig.~\ref{fig:effect_bulkvisc_vtk}.}
	\label{fig:effect_bulkvisc_spectrum}
\end{figure}

The density field at the end of the simulation is shown in Fig.~\ref{fig:effect_bulkvisc_vtk} where the case $\Lambda_b=1$, resembling a standard TRT method, is compared to cases for $\Lambda_b=10$ and $\Lambda_b=100$ throughout the whole domain.
It is clearly visible that the oscillations get damped when increasing $\Lambda_b$ and even vanish almost completely for $\Lambda_b=100$.
This is especially true for the density spikes right along the surface of the sphere.
At last, the adaptive approach TRT+aB is taken where $\Lambda_b=100$ is only set inside a spherical shell with a width of two fluid cells around the sphere.
The observable damping effect is comparable to TRT+B($\Lambda_b=10$), but additionally, the near-surface density spikes are smoothed out more and thus similar to TRT+B($\Lambda_b=100$).

A more detailed analysis of the damping properties can be gained from the spectrum of the density signal at the probe location, shown in Fig.~\ref{fig:effect_bulkvisc_spectrum}.
The low-frequency parts remain almost unchanged in all cases, whereas the damping properties can be observed in the middle to higher frequency regions.
There, the case TRT+B($\Lambda_b=100$) exhibits almost no contributions.
The direct comparison of TRT+B($\Lambda_b=10$) with TRT+aB($\Lambda_b=100$) shows a generally very similar behavior, with a better high-frequency damping for the former and a better mid-frequency damping for the latter one.

Concluding, this test shows that our proposed modifications of the LBM collision operator are a viable tool to reduce the density fluctuations introduced by the mapping update.
It also shows that increasing the bulk viscosity only in the vicinity of the fluctuation source can be sufficient to reduce the oscillations globally.
Finally, we note that overall density fluctuations in this test case are very small and below $0.01\%$ of the average density.
Depending on the parameterization, however, they could be larger and then also have a clear influence on the trajectories of other particles and on the overall stability of the simulation.

\subsection{Force on infinite array of fixed spheres in Stokes flow}
\label{sec:StokesDrag}

\subsubsection{Background}

As a first test to establish the physical correctness of the approach, specifically the fluid solver and the computation of the fluid-particle interaction force, we simulate Stokes flow around an infinite and steady array of spheres.
Once converged, we can evaluate the force acting on the sphere and compare it against existing quasi-analytical solutions~\cite{sangani1982}.
Due to its simplicity and the existence of a quasi-analytical solution, it is a well-studied setup, both for classical DNS approaches~\cite{tang2014} as well as for the LBM \cite{pan2006,khirevich2015,rettinger2017,khirevich2018}.
For LBM simulations, this is a particularly important test as some collision operators suffer from an undesired dependence of the drag force on the relaxation rate $s_\nu$, which affects the simulated boundary location~\cite{pan2006}.
As mentioned in Sec.~\ref{sec:lbm}, the TRT collision operator has been designed specifically to overcome this issue.
We will thus use this test case to ensure the correctness of our implementation and, furthermore, to show that the introduction of the third relaxation time via $\Lambda_b$ in the TRT+B and TRT+aB collision model does not change this important property.

\subsubsection{Description}

\begin{figure}[t]
	\centering
	\begin{subfigure}[b]{0.44\textwidth}
		\centering
		\includegraphics[width=\textwidth]{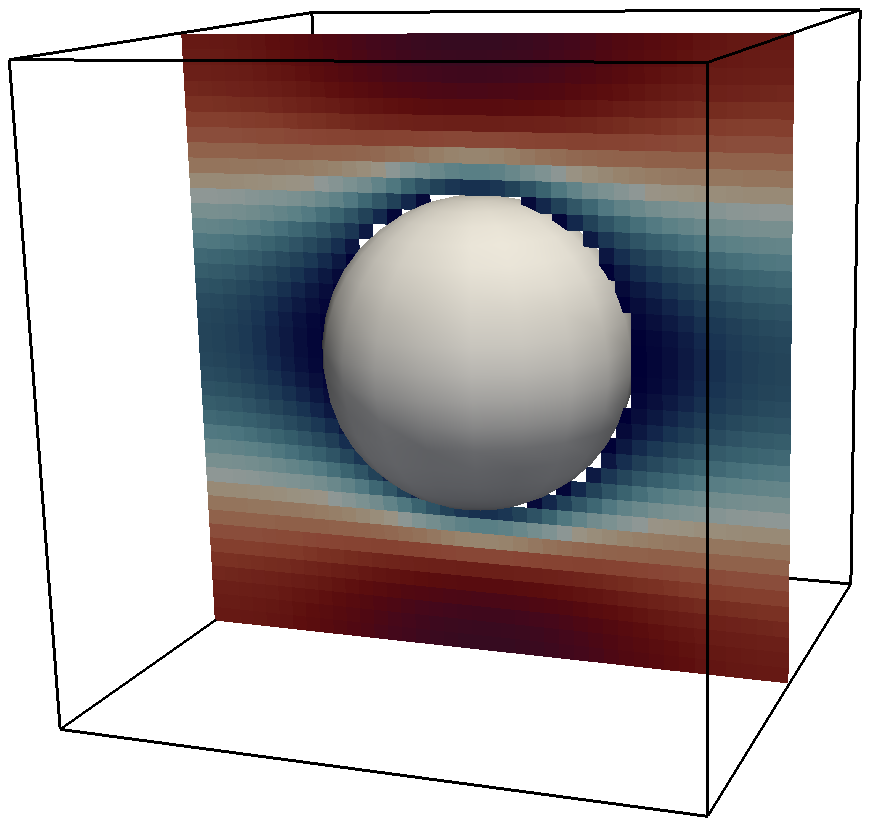}
		\caption{Visualization of the flow field.}
		\label{fig:setup_force_on_sphere_stokes}
	\end{subfigure}~
\begin{subfigure}[b]{0.55\textwidth}
		\centering	
	\input{figures/drag_force_on_sphere.pgf}
	\caption{Relative error of the total force.}
	\label{fig:force_on_sphere_stokes}
\end{subfigure}
\caption{Setup and result of force evaluation on infinite array of fixed spheres in Stokes flow.}
\end{figure}

We use the setup from Ref.~\citenum{rettinger2017}, i.e., a fully periodic domain of size $L\times L \times L = [2\times 2\times 2]D_p$, with a single sphere of diameter $D_p=20$ located in the center of the domain.
The flow is driven by an external force $\mathbf{a}=(a,0,0)^\top$ with $a=10^{-8}$ to ensure Stokes flow conditions for all viscosities.
A visualization of the setup and the resulting flow is shown in Fig.~\ref{fig:setup_force_on_sphere_stokes}.
The fluid-particle interaction force, Eq.~\eqref{eq:interaction_force}, together with the buoyancy force, $\mathbf{F}_p^{buoy} = \tfrac{\pi}{6}D_p^3\mathbf{a}$, then yields the total normalized force in forcing direction
\begin{equation}
C = \frac{\tfrac{1}{|\mathbf{a}|}\mathbf{a} \cdot (\mathbf{F}_p^{fp} + \mathbf{F}_p^{buoy})}{ 3\pi \mu_f D_p \bar{u}},
\end{equation}
with the average fluid velocity 
\begin{equation}
\bar{u} = \tfrac{1}{|\mathbf{a}|L^3}\mathbf{a} \cdot \sum_{\mathbf{x}} \mathbf{u}_f (\mathbf{x}).
\end{equation}
For that case, the reference value is given as $C^{ref} = 2.842$~\cite{sangani1982}.

\subsubsection{Results and Discussion}

To study the effect of the viscosity value onto the total normalized force, we use different values for $s_\nu \in \left\{\tfrac{1}{0.55}, \tfrac{1}{0.6}, \tfrac{1}{0.7}, \tfrac{1}{0.8}, \tfrac{1}{0.9}, 1, \tfrac{2}{3}\right\}$ and check the simulated force against the reference value for different LBM collision models.
The outcome is shown in Fig.~\ref{fig:force_on_sphere_stokes} as the relative error of the force.
As expected, the SRT collision model yields a force that depends on the choice of the relaxation rate.
For low viscosities, this error stays below 2\% whereas it grows steadily for larger viscosity values.
This is thus particularly problematic for low Reynolds number flows in porous media where the observed permeability can be influenced significantly by this effect~\cite{pan2006}.
All other here tested collision models, however, show no dependence on the viscosity and can accurately recover the reference solution.
This demonstrates that adding the control over the bulk viscosity via the parameter $\Lambda_b$, either globally or only in the region around the sphere, maintains the desired TRT property and its accuracy. 


\subsection{Drag and lift forces on spherical particle in shear flow}

\subsubsection{Background}

In the next step, we consider a steady sphere that is placed close to the resting plane in a uniform shear flow setup.
This way, we validate the capability of our approach to predict drag and lift forces on this sphere for different Reynolds numbers, thus extending the flow regime of the previous test.
For such a setup, empirical drag and lift correlations have been obtained by accurate spectral element simulations \cite{zeng2009}, which will act as a reference.
Besides the shear rate, the forces depend on the sphere's distance from the bottom wall and we can thus investigate the behavior of our approach for cases where the distance between two objects is smaller than a fluid cell.
Many sediment transport processes feature shear flows where the interplay of drag and lift forces is crucial for the onset of particle motion and the transport mode \cite{agudo2017}.
Thus, this test case validates the applicability of a numerical simulation for such setups.

\subsubsection{Description}

\begin{figure}[t]
	\centering
	\includegraphics[width=0.8\textwidth]{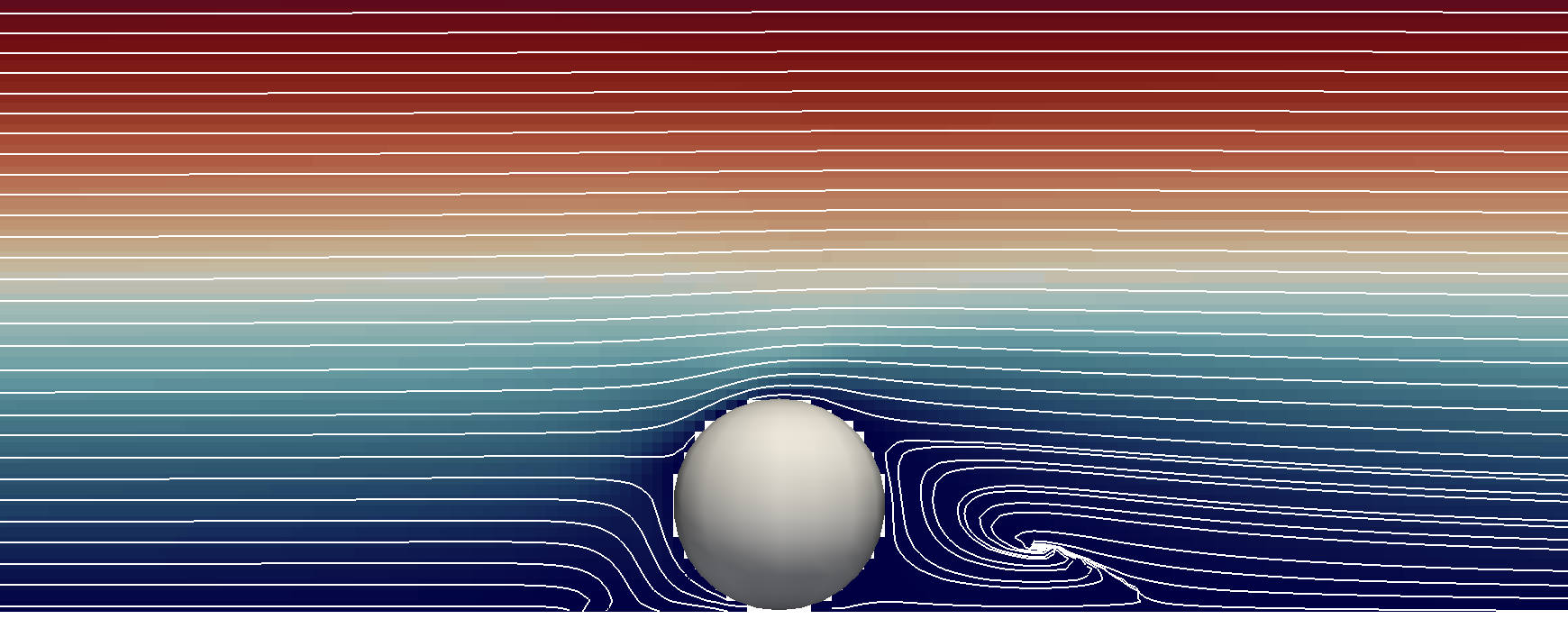}
	\caption{Visualization of the flow field around the stationary sphere for $Re_s=25$ and $L=0.505$.}
	\label{fig:setup_forcesOnSphereNearPlane}
\end{figure}

The setup features a domain of size $[48\times16\times8]D_p$, which is periodic in $x-$ and $y-$direction.
A stationary sphere with diameter $D_p=20$ is placed at position $(24,8,L)D_p$, i.e., horizontally centered and with a given dimensionless distance $L$ from the bottom wall.
The top wall is moving with a constant velocity $\mathbf{u}_w = (u_w,0,0)^\top$ while the bottom wall is at rest.
A similar setup has also been used in Ref.~\citenum{lee2010} to validate a classical DNS approach.
Here, we use $u_w =0.1$.
The flow is characterized by the shear Reynolds number $Re_s = G D_p^2/ \nu_f$, where the shear rate of the undisturbed flow is given as $G=u_w / H$ and the domain height $H=8 D_p$.
The resulting shear flow induces drag and lift forces onto the sphere, which depend on $Re_s$ and the sphere's distance to the bottom wall \cite{zeng2009}.
Those are evaluated as the hydrodynamic forces acting on the sphere in $x-$ and $z-$ direction, respectively, and then normalized to obtain the drag and lift coefficients:
\begin{equation}
C_d = \frac{F_{p,x}^{fp}}{\tfrac{\pi}{8}\rho_f G^2 L^2 D_p^4},\quad C_L = \frac{F_{p,z}^{fp}}{\tfrac{\pi}{8}\rho_f G^2 L^2 D_p^4}
\end{equation}
We initialize the domain with the linear shear profile and run the simulation until convergence of drag and lift is observed.
A magnified visualization of the flow field around the sphere for the case $Re_s=25$ and $L=0.505$ is shown in Fig.~\ref{fig:setup_forcesOnSphereNearPlane}.

\subsubsection{Results and Discussion}

We evaluate drag and lift coefficients for two different distances, $L = 0.505$ and $1$, and two flow configurations, $Re_s = 1$ and $25$.
The results are reported in Tabs.~\ref{tab:drag_lift_sphere_Re1} and \ref{tab:drag_lift_sphere_Re25}.
For $Re_s=1$, an excellent agreement of the simulated drag and lift values to the proposed correlations from \cite{zeng2009} can be seen.
This is especially true for $L=0.505$, where it has to be noted that with the chosen $D_p=20$, the resolution is too coarse to resolve the flow in the gap between the sphere's surface and the plane.
Also, there is no noticeable difference between the three different LBM collision model variants, demonstrating once more the applicability of our suggested improvements.

In principle, a similar conclusion can be drawn from the results for $Re_s=25$.
The errors there, especially of the lift coefficient for $L=1$, are slightly larger than for the low $Re_s$ case.
This, however, can be attributed to the fact that the used reference values originate from fits to simulated data~\cite{zeng2009}.
Those correlations capture the underlying data with deviations of around 5\% for $C_D$ and up to 16\% for $C_L$~\cite{zeng2009}.
For the collision model TRT+B ($\Lambda_b=100$), no results could be obtained since the simulations became unstable.
This is unexpected as the modifications were originally made to stabilize the simulation.
Further studies of an empty sheared channel, where the same issues could be observed, led to the conclusion that this variant might be prone to inaccuracies introduced when initializing the PDF values using only density and velocity information, neglecting higher-order components.
Additionally, the particular choice of $\Lambda$, from Eq.~\eqref{eq:MagicParameter}, can stabilize the simulation but at the same time reduces the overall accuracy, as also noted by Ref.~\citenum{khirevich2015}.
A detailed study would, however, be out of scope of this paper and we, therefore, use the variant TRT+aB ($\Lambda_b=100$) for the remainder of this work.

\begin{table}[t]
	\caption{Drag and lift coefficients for a sphere close to a wall and $Re_s = 1$.}
	\label{tab:drag_lift_sphere_Re1}
	\centering
	\begin{tabular}{lc||l|cc|cc|cc|}
		  &       &        & \multicolumn{2}{|c|}{TRT ($\Lambda_b=1$)}& \multicolumn{2}{|c|}{TRT+B ($\Lambda_b=100$)} & \multicolumn{2}{|c|}{TRT+aB ($\Lambda_b=100$)} \\
		  &       & Ref \cite{zeng2009}  & Sim   & rel. Err (\%) & Sim   & rel. Err (\%) & Sim   & rel. Err (\%)\\\hline
$L=0.505$ & $C_D$ & 44.88                & 43.98 & 2.00          & 44.04 & 1.87          & 44.00 & 1.97 \\
		  & $C_L$ & 3.54                 & 3.43  & 2.95          & 3.44  & 2.78          & 3.43  & 2.91 \\ \hline  
$L=1$     & $C_D$ & 35.77                & 34.43 & 3.74          & 34.45 & 3.69          & 34.44 & 3.72 \\
		  & $C_L$ & 2.32                 & 2.30  & 0.87          & 2.30  & 0.79          & 2.30  & 0.83 \\ \hline 
	\end{tabular}
\end{table}

\begin{table}[t]
	\caption{Drag and lift coefficients for a sphere close to a wall and $Re_s = 25$.}
	\label{tab:drag_lift_sphere_Re25}
	\centering
	\begin{tabular}{lc||l|cc|cc|cc|}
		  &       &        & \multicolumn{2}{|c|}{TRT ($\Lambda_b=1$)}& \multicolumn{2}{|c|}{TRT+B ($\Lambda_b=100$)} & \multicolumn{2}{|c|}{TRT+aB ($\Lambda_b=100$)} \\
		  &       & Ref \cite{zeng2009}  & Sim   & rel. Err (\%) & Sim   & rel. Err (\%) & Sim   & rel. Err (\%)\\\hline
$L=0.505$ & $C_D$ & 3.53                 & 3.70 &  4.62          & -     & -             & 3.70  & 4.67 \\
		  & $C_L$ & 0.88                 & 0.91 &  3.46          & -     & -             & 0.92  & 3.90 \\ \hline  
$L=1$     & $C_D$ & 2.81                 & 2.78 &  1.09          & -     & -             & 2.78  & 1.06 \\
          & $C_L$ & 0.45                 & 0.41 &  8.05          & -     & -             & 0.41  & 8.04 \\ \hline 
	\end{tabular}
\end{table}

\section{Particle interaction based on lubrication}

\label{sec:lubrication_interactions}

\subsection{Model for unresolved lubrication interactions}
\label{sec:lubrication_model}

When two particles approach each other and are getting close, the fluid inside the forming gap between the two surfaces is squeezed out.
Depending on the flow conditions, this effect can lead to large resistances, given as lubrication forces and torques opposing the relative motion.
These effects influence the particle interaction behavior significantly.
In order to accurately capture them with a fluid-particle coupling algorithm, like the one from Sec.~\ref{sec:coupling}, a very fine resolution of the gap would be required, which is computationally not feasible~\cite{ladd2001}. 
For that reason, it is common to introduce lubrication correction models that aim to compensate the unresolved lubrication interactions~\cite{ladd2001,nguyen2002,breugem2010,costa2015,biegert2017}.
As a result, the total hydrodynamic force $\boldsymbol{F}_{p,i}^{hyd}$ and torque $\boldsymbol{T}_{p,i}^{hyd}$ acting on a particle $i$ consist of a part that is fully resolved by the fluid-particle coupling method from Sec.~\ref{sec:coupling}, and the lubrication correction, resulting in
\begin{align}
\boldsymbol{F}_{p,i}^{hyd} &= \boldsymbol{F}_{p,i}^{fp} +  \boldsymbol{F}_{p,i}^{\text{lub,cor}}, \label{eq:hydrodynamic_force}\\
\boldsymbol{T}_{p,i}^{hyd} &= \boldsymbol{T}_{p,i}^{fp} +  \boldsymbol{T}_{p,i}^{\text{lub,cor}}. \label{eq:hydrodynamic_torque}
\end{align}

These corrections are typically the leading order terms of the analytical lubrication forces, derived for Stokes flow  and thus split into normal and tangential contributions~\cite{brenner1961,jeffrey1984}
\begin{align}
\boldsymbol{F}_{p,i}^{\text{lub,cor}} &= \sum_{j \neq i}\left(\boldsymbol{F}_{ij,n}^{\text{lub,cor}} + \boldsymbol{F}_{ij,tt}^{\text{lub,cor}} + \boldsymbol{F}_{ij,tr}^{\text{lub,cor}}\right),\label{eq:lubrication_correction_force_total}\\
\boldsymbol{T}_{p,j}^{\text{lub,cor}} &= \sum_{j \neq i}\left(\boldsymbol{T}_{ij,tt}^{\text{lub,cor}} + \boldsymbol{T}_{ij,tr}^{\text{lub,cor}}\right). \label{eq:lubrication_correction_torque_total}
\end{align}

Assuming two particles $i$ and $j$ with different radii, $R_{p,i}$ and $R_{p,j}$, we define the radius ratio $\kappa_r = R_{p,j}/R_{p,i}$.
Then, the lubrication correction in normal direction that is applied on particle $i$ can be written as~\cite{jeffrey1984,ladd2001,nguyen2002}
\begin{equation}
\boldsymbol{F}_{ij,n}^{\text{lub,cor}} =
-6\pi\mu_f R_{p,i}^2  \lambda(\delta_n^\text{lub},\delta_{n,\text{cut}}^{\text{lub}}) \frac{\kappa_r^2}{(1+\kappa_r)^2}\left(\frac{1}{\delta_n^\text{lub}} - \frac{1}{\delta_{n,\text{cut}}^{\text{lub}}} \right) \boldsymbol{u}_{ij,n},
\label{eq:lubrication_correction_force_normal}
\end{equation}
with the dynamic fluid viscosity $\mu_f$, the normal relative particle velocity according to the definitions in \ref{app:particleDefinitions}, and the gap size
\begin{equation}
\delta_n^\text{lub} = \max(\delta_{ij,n},\delta_{n,\text{min}}^{\text{lub}}). \label{eq:lubrication_correction_gap_size}
\end{equation}
The parameter $\delta_{n,\text{min}}^{\text{lub}}$ effectively saturates the lubrication force for very small gaps by avoiding a division by zero and its value is often related physically to surface asperities~\cite{izard2014}.
We will study the effect of this parameter in Sec.~\ref{sec:sphereWallCollision} in more detail. 
The step function 
\begin{equation}
\lambda(\delta_n,\delta_{\text{cut}}^{\text{lub}}) = 
\begin{cases}
1,&\quad 0 < \delta_n < \delta_{\text{cut}}^{\text{lub}}, \\
0,&\quad \text{otherwise,}
\end{cases}
\label{eq:lubrication_correction_step_function}
\end{equation}
is applied to switch off the correction during collisions and when the normal surface distance $\delta_n$ is larger than a cut-off value $\delta_{\text{cut}}^{\text{lub}}$.
This introduces another parameter $\delta_{n,\text{cut}}^{\text{lub}}$, i.e., the cut-off distance for the lubrication correction in normal direction.
This parameter is the distance until which the fluid-particle coupling algorithm can yield accurate hydrodynamic interaction information and will be determined in Sec.~\ref{sec:test_lubricationcorrection}.

In tangential direction, the lubrication correction due to a tangential translational velocity difference can be formulated as \cite{jeffrey1984, simeonov2012}
\begin{align}
\boldsymbol{F}_{ij,tt}^{\text{lub,cor}} &= 6 \pi \mu_f R_{p,i}  \lambda(\delta_n^\text{lub},\delta_{tt,\text{cut}}^{\text{lub}})  \frac{4 \kappa_r(2+\kappa_r + 2 \kappa_r^2)}{15(1+\kappa_r)^3}\ln\left(\frac{\delta_n^\text{lub}}{\delta_{tt,\text{cut}}^{\text{lub}}}\right) \boldsymbol{u}_{ij,t} \label{eq:lubrication_correction_force_tangential_trans}\\
\boldsymbol{T}_{ij,tt}^{\text{lub,cor}} &= 8 \pi \mu_f R_{p,i}^2  \lambda(\delta_n^\text{lub},\delta_{tt,\text{cut}}^{\text{lub}})  \frac{\kappa_r (4+\kappa_r)}{10(1+\kappa_r)^2} \ln\left(\frac{\delta_n^\text{lub}}{\delta_{tt,\text{cut}}^{\text{lub}}}\right) (\boldsymbol{u}_{ij} \times \boldsymbol{n}_{ij})
\label{eq:lubrication_correction_torque_tangential_trans}
\end{align}
and for a relative angular velocity $\boldsymbol{\omega}_{ij} = \boldsymbol{\omega}_i + \boldsymbol{\omega}_j$ around an axis perpendicular to $\boldsymbol{n}_{ij}$
\begin{align}
\boldsymbol{F}_{ij,tr}^{\text{lub,cor}} &= -6 \pi \mu_f R_{p,i}^2  \lambda(\delta_n^\text{lub},\delta_{tr,\text{cut}}^{\text{lub}})  \frac{2\kappa_r^2}{15 (1+\kappa_r)^2} \ln\left(\frac{\delta_n^\text{lub}}{\delta_{tr,\text{cut}}^{\text{lub}}}\right) \left(\boldsymbol{\omega}_{ij} + \frac{4}{\kappa_r}\boldsymbol{\omega}_i + 4\kappa_r \boldsymbol{\omega}_j\right) \times \boldsymbol{n}_{ij} \label{eq:lubrication_correction_force_tangential_rot} \\
\boldsymbol{T}_{ij,tr}^{\text{lub,cor}} &= 8 \pi \mu_f R_{p,i}^3  \lambda(\delta_n^\text{lub},\delta_{tr,\text{cut}}^{\text{lub}}) \frac{2\kappa_r}{5(1+\kappa_r)}\ln\left(\frac{\delta_n^\text{lub}}{\delta_{tr,\text{cut}}^{\text{lub}}}\right)\left(\boldsymbol{\omega}_i + \frac{\kappa_r}{4}\boldsymbol{\omega}_j - \left(\left(\boldsymbol{\omega}_i + \frac{\kappa_r}{4}\boldsymbol{\omega}_j\right) \cdot \boldsymbol{n}_{ij}\right)\boldsymbol{n}_{ij}\right)
\label{eq:lubrication_correction_torque_tangential_rot}
\end{align}
As for the normal direction, cut-off distances $\delta_{tt,\text{cut}}^{\text{lub}}$ and $\delta_{tr,\text{cut}}^{\text{lub}}$ are introduced as parameters~\cite{nguyen2002,janoschek2013} and have to be determined adequately.
Following Ref.~\citenum{simeonov2012}, we are excluding the torque due to a relative rotation about the line of centers which becomes negligibly small for small gap sizes.
The corresponding lubrication corrections for sphere-wall interactions are obtained by assuming $\kappa_r\rightarrow\infty$~\cite{nguyen2002,simeonov2012}.

\subsection{Determination of the lubrication correction cut-off distances}

\label{sec:test_lubricationcorrection}

\subsubsection{Background}

The hydrodynamic interactions of two spheres with relative translational or angular velocities have been studied in detail for low Reynolds number flows and analytical approximations exist~\cite{jeffrey1984}.
It has been noted in coupled fluid-particle simulations with classical DNS approaches \cite{breugem2010, costa2015} or LBM \cite{ladd2001,nguyen2002, janoschek2013,bartuschat2015} that the simulation predictions of the hydrodynamic interaction for this case break down when the gap is smaller than a certain gap width.
In those cases, the lubrication correction models from the previous section have to be applied to compensate for these otherwise missing interactions.
For that purpose, it is crucial to determine these gap widths for the applied coupling method as those values are taken as the cut-off distances $\delta_{n,\text{cut}}^{\text{lub}}, \delta_{tt,\text{cut}}^{\text{lub}},$ and $\delta_{tr,\text{cut}}^{\text{lub}}$ in Eq.~\eqref{eq:lubrication_correction_step_function}.
This procedure effectively avoids that the corrections are applied in cases where the coupling method still yields the correct interactions, which would otherwise result in their overestimation.
Those cut-off distances generally have to be found by appropriate simulation tests, like the following one.

\subsubsection{Description}

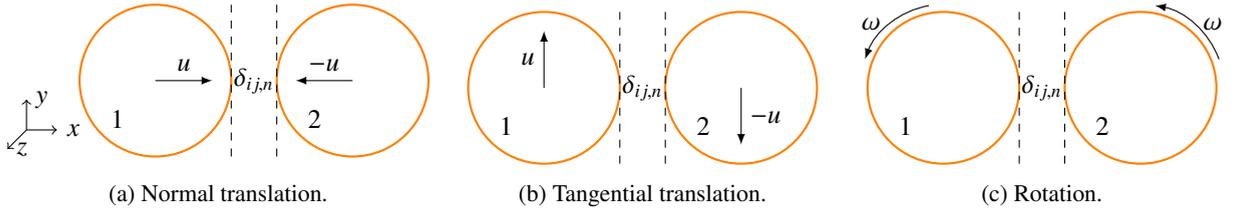
\begin{figure}[t]
	\centering
\begin{subfigure}[t]{0.35\textwidth}
	\centering
\begin{tikzpicture}[]
\draw[->] (-1.7,-0.65) -- ++(0.4,0) node[pos=1,right]{$x$};
\draw[->] (-1.7,-0.65) -- ++(0,0.4) node[pos=1,right]{$y$};
\draw[->] (-1.7,-0.65) -- ++(-0.25,-0.25) node[pos=1,right]{$z$};
\coordinate[] (xpi) at (0,0);
\coordinate[label=center:1] (number1) at ($(xpi)-(0.5,0.5)$ );
\coordinate[] (xpj) at (2.6,0);
\coordinate[label=center:2] (number2) at ($(xpj)-(0.5,0.5)$ );
\draw[orange,thick] (xpi) circle (1);
\draw[orange,thick] (xpj) circle (1);
\draw[-latex] (xpi) -- ++(0.75,0) node[pos=0.5,above]{$u$};
\draw[-latex] (xpj) -- ++(-0.75,0) node[pos=0.5,above]{$-u$};
\draw[dashed] (1,-1) -- ++(0,2);
\draw[dashed] (1.6,-1) -- ++(0,2);
\coordinate[label=center:$\delta_{ij,n}$] (label) at ($0.5*(xpj)+0.5*(xpi)$ );
\end{tikzpicture}
\caption{Normal translation.}
\end{subfigure}~
\begin{subfigure}[t]{0.31\textwidth}
	\centering
\begin{tikzpicture}[]
\coordinate[] (xpi) at (0,0);
\coordinate[label=center:1] (number1) at ($(xpi)-(0.5,0.5)$ );
\coordinate[] (xpj) at (2.6,0);
\coordinate[label=center:2] (number2) at ($(xpj)-(0.5,0.5)$ );
\draw[orange,thick] (xpi) circle (1);
\draw[orange,thick] (xpj) circle (1);
\draw[-latex] (xpi) -- ++(0,0.75) node[pos=0.5,left]{$u$};
\draw[-latex] (xpj) -- ++(0,-0.75) node[pos=0.5,right]{$-u$};
\draw[dashed] (1,-1) -- ++(0,2);
\draw[dashed] (1.6,-1) -- ++(0,2);
\coordinate[label=center:$\delta_{ij,n}$] (label) at ($0.5*(xpj)+0.5*(xpi)$ );
\end{tikzpicture}
\caption{Tangential translation.}
\end{subfigure}~
\begin{subfigure}[t]{0.31\textwidth}
	\centering
\begin{tikzpicture}[]
\coordinate[] (xpi) at (0,0);
\coordinate[label=center:1] (number1) at ($(xpi)-(0.5,0.5)$ );
\coordinate[] (xpj) at (2.6,0);
\coordinate[label=center:2] (number2) at ($(xpj)-(0.5,0.5)$ );
\draw[orange,thick] (xpi) circle (1);
\draw[orange,thick] (xpj) circle (1);
\draw[-latex] ($(xpi) + (100:1.1)$) arc (100:160:1.1) node[pos=0.5,left]{$\omega$};
\draw[-latex] ($(xpj) + (20:1.1)$) arc (20:80:1.1) node[pos=0.5,right]{$\omega$};
\draw[dashed] (1,-1) -- ++(0,2);
\draw[dashed] (1.6,-1) -- ++(0,2);
\coordinate[label=center:$\delta_{ij,n}$] (label) at ($0.5*(xpj)+0.5*(xpi)$ );
\end{tikzpicture}
\caption{Rotation.}
\end{subfigure}
\caption{Test setups for the determination of the cut-off distance of the lubrication correction model.}
\label{fig:sketch_lubrication}
\end{figure}

Two spheres are placed next to each other with a given surface distance $\delta_{ij,n}$ submerged in a fluid.
By imposing a certain relative velocity, we can observe the resulting hydrodynamic interaction force and torque, and compare it to the analytical predictions \cite{jeffrey1984}.
We are then investigating up to which gap size our fluid-particle coupling approach yields reliable force information and when lubrication corrections become necessary, which then determines the cut-off distance for the lubrication correction.
Since we account for normal and tangential lubrication forces, whereas the latter can originate from relative translating or rotating motion, we have to determine three different cut-off distances, namely $\delta_{n,\text{cut}}^{\text{lub}}$ from Eq.~\eqref{eq:lubrication_correction_force_normal}, $\delta_{tt,\text{cut}}^{\text{lub}}$ from Eqs.~\eqref{eq:lubrication_correction_force_tangential_trans} and \eqref{eq:lubrication_correction_torque_tangential_trans}, and $\delta_{tr,\text{cut}}^{\text{lub}}$ from Eqs.~\eqref{eq:lubrication_correction_force_tangential_rot} and \eqref{eq:lubrication_correction_torque_tangential_rot}.
As these contributions can be treated independently of each other, we consider three different setups that can be seen in Fig.~\ref{fig:sketch_lubrication}.

The domain is of size $[12\times12\times12]D_p$ and fully periodic.
The first sphere (1) is placed inside the center of the domain, whereas the second one (2) is offset by $(D_p+\delta_{ij,n},0,0)$.
Depending on the case, we set a translational velocity $u$ or a rotational velocity $\omega=2u/D_p$ onto the spatially fixed spheres, see Fig.~\ref{fig:sketch_lubrication}, similar to the setup used in Ref.~\citenum{breugem2010}.
We use $\nu_f=1/6$ and $Re=D_p u / \nu_f = 0.01$, and run the simulation until convergence of the forces and torques.
We carry out the simulation for two different diameters, $D_p = 10$ and $20$, to investigate whether the cut-off distance should be regarded as a function of the diameter.

Since the source of the inaccurate prediction is due to insufficient grid resolution inside the gap between the surfaces, it can be assumed that the specific placement of the particles on the grid might additionally influence the force values.
Due to the explicit particle mapping, there might be cases where there is still a fluid cell inside the gap, whereas there might be none if the spheres were located differently on the grid.
To determine the cut-off distances in a general way, we follow the approach from Ref.~\citenum{janoschek2013} and carry out each single simulation setup 15 times with each time a different random offset applied to both, sphere 1's and sphere 2's, positions.
Afterwards, we average the resulting force and torque over these realizations.

\subsubsection{Results and Discussion}

\begin{figure}
	\centering	
	\input{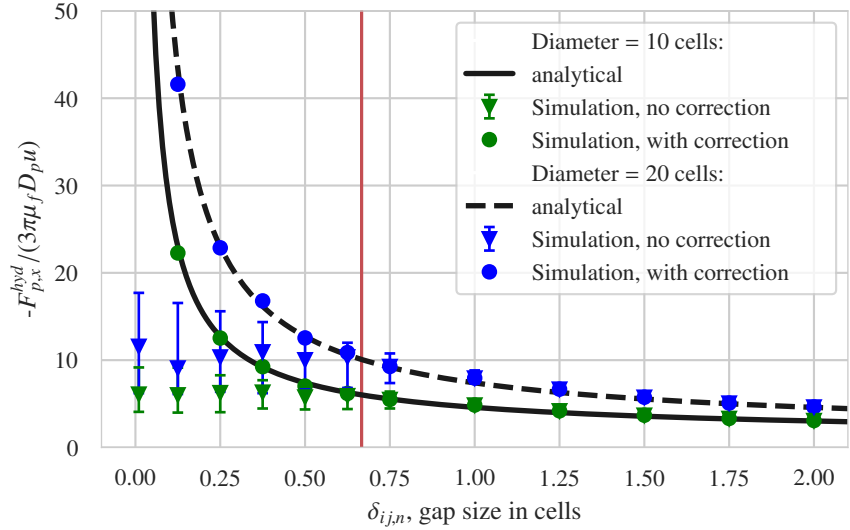}
	\caption{Hydrodynamic force on sphere 1 as a function of the gap size for the case of two normally translating spheres for two different diameters. The respective analytical near-field predictions from Ref.~\citenum{jeffrey1984} are given as black lines. The simulation results, averaged over the 15 realizations, are shown with and without lubrication correction. For the latter one, error bars are included that indicate the minimum and maximum value of the realizations.}
	\label{fig:lubrication_sphsph_normal_force_test}
\end{figure}

\begin{figure}[t]
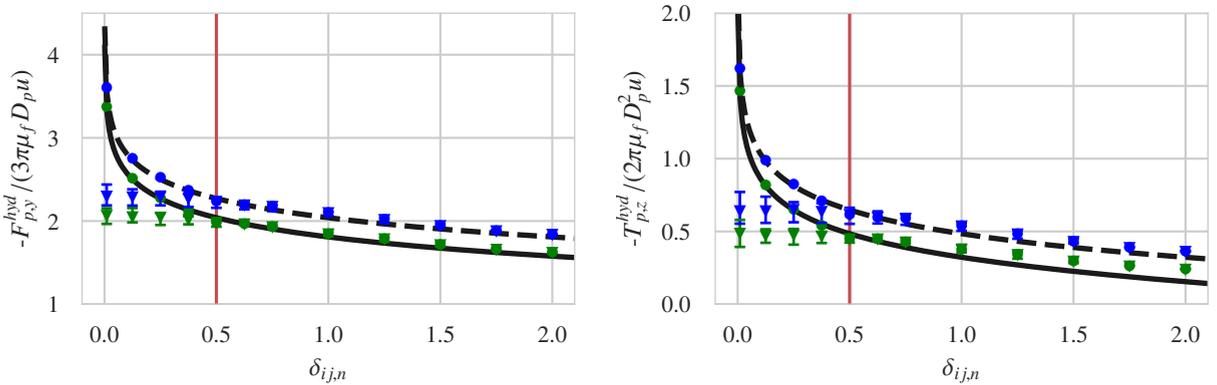

	\centering
	\begin{subfigure}[b]{0.49\textwidth}
	\centering	
	\input{figures/lubrication_sphsph_tangentialtranslation_force.pgf}
	\label{fig:lubrication_sphsph_tangentialtranslationl_force_test}
	\end{subfigure}~
	\begin{subfigure}[b]{0.49\textwidth}
	\centering	
	\input{figures/lubrication_sphsph_tangentialtranslation_torque.pgf}
	\label{fig:lubrication_sphsph_tangentialtranslation_torque_test}
	\end{subfigure}
	\caption{Hydrodynamic force (left) and (torque) on sphere 1 for the case of two tangentially translating spheres. Colors and style as in Fig.~\ref{fig:lubrication_sphsph_normal_force_test}}
	\label{fig:lubrication_sphsph_tangentialtranslation_test}
\end{figure}

\begin{figure}[t]
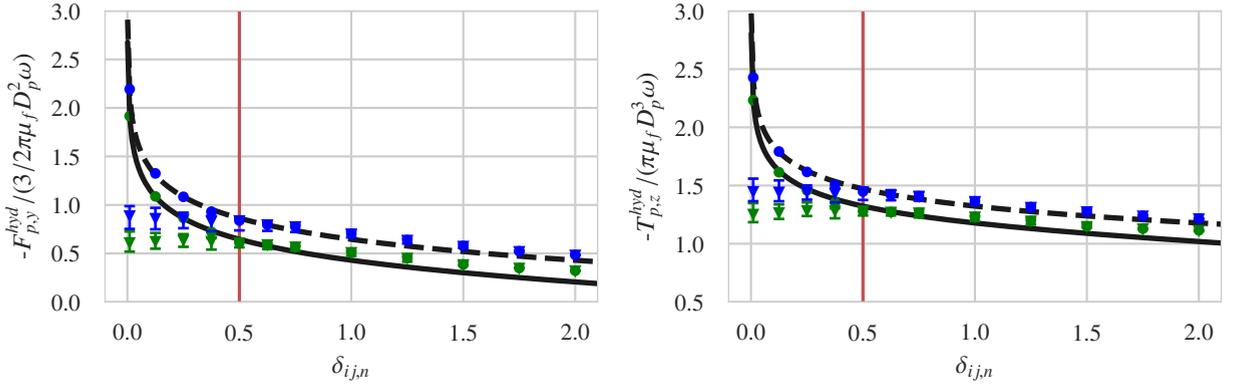

	\centering
	\begin{subfigure}[b]{0.49\textwidth}
		\centering	
		\input{figures/lubrication_sphsph_tangentialrotation_force.pgf}
		\label{fig:lubrication_sphsph_tangentialrotation_force_test}
	\end{subfigure}~
	\begin{subfigure}[b]{0.49\textwidth}
		\centering	
		\input{figures/lubrication_sphsph_tangentialrotation_torque.pgf}
		\label{fig:lubrication_sphsph_tangentialrotation_torque_test}
	\end{subfigure}
	\caption{Hydrodynamic force (left) and (torque) on sphere 1 for the case of two rotating spheres. Colors and style as in Fig.~\ref{fig:lubrication_sphsph_normal_force_test}}
	\label{fig:lubrication_sphsph_tangentialrotation_test}
\end{figure}

The normalized force acting on sphere 1 for the case of the normally translating spheres is shown in Fig.~\ref{fig:lubrication_sphsph_normal_force_test} over the gap size.
Additionally, the analytical prediction from Ref.~\citenum{jeffrey1984} allows for continuous evaluation and is given in black.
We note that by applying a physical scaling to the $x$-axis with $D_p$, the analytical curves would collapse into a single one, as expected.
We can observe that the simulation without lubrication correction accurately predicts the interaction force for larger gap sizes.
For gap sizes smaller than around $2/3$ of the cell size, indicated by the red vertical line, the simulations can not follow the predicted strong increase and instead remain almost constant.
The hydrodynamic force would thus be considerably underestimated.
This behavior is the same for $D_p=10$ and $20$.
When adding the lubrication correction in the normal direction with the determined value of $\delta_{n,\text{cut}}^{\text{lub}} = 2/3\Delta x$, this deficit can be cured and the simulation results match the predictions very well.

The results for tangentially translating and rotating spheres can be seen in Fig.~\ref{fig:lubrication_sphsph_tangentialtranslation_test} and Fig.~\ref{fig:lubrication_sphsph_tangentialrotation_test}, respectively.
In all cases, it can be seen that the simulated force and torque follow the predictions accurately until a certain gap size is reached from whereon the corresponding lubrication corrections are necessary.
For these cases, this limiting gap size is given as $\tfrac{1}{2}$ of the cell size, independent of force or torque.

Summarizing, this test allows to determine the cut-off distances for the lubrication correction concisely.
We will use 
\begin{equation}
\delta_{n,\text{cut}}^{\text{lub}} = \tfrac{2}{3}\Delta x, \delta_{tt,\text{cut}}^{\text{lub}} = \tfrac{1}{2}\Delta x, \delta_{tr,\text{cut}}^{\text{lub}} = \tfrac{1}{2}\Delta x 
\label{eq:lubricationCutOff}
\end{equation}
for the remainder of this work.
This is similar to the values found by others using LBM, but with different LBM collision models and boundary conditions~\cite{nguyen2002,janoschek2013}.
We thus do not follow Refs.~\citenum{breugem2010,costa2015}, who suggest a dependence of the cut off distance on the diameter since, as we have shown, these cut-off distances are a numerical parameter without a physical counterpart.

As can be seen when comparing the force and torque magnitudes of the normal and tangential components in Figs.~\ref{fig:lubrication_sphsph_normal_force_test},\ref{fig:lubrication_sphsph_tangentialtranslation_test} and \ref{fig:lubrication_sphsph_tangentialrotation_test}, the tangential interactions are much weaker than the normal one and, consequently, less critical.
This is the reason why the tangential contributions are often not considered in the literature~\cite{kempe2012,costa2015,biegert2017}.
This might be justified in some cases but could lead to an underestimation of tangential interaction e.g., in dense particulate systems at low Reynolds numbers, and thus should be decided on a case by case basis.

\section{Particle interaction based on collisions}

\label{sec:collisional_interactions}

\subsection{Particle dynamics with the discrete element method}
\label{sec:rpd}

In this work, we employ a discrete element method (DEM) to describe the collision behavior of particles with other particles or walls.
After all forces and torques acting on the particles are known, their position and velocity are updated by a Velocity-Verlet integrator.
As the simulations in this work only feature spherical particles, we will restrict the explanations to spheres, noting that the underlying ideas can readily be applied to non-spherical particles.
The used notation is given in \ref{app:particleDefinitions}.

\subsubsection{Equations of particle motion}

The motion of each individual particle is generally described by 
\begin{align}
m_{p,i} \frac{\text{d} \boldsymbol{u}_{p,i}}{\text{d} t} &=  \boldsymbol{F}_{p,i} = \boldsymbol{F}_{p,i}^{col} + \boldsymbol{F}_{p,i}^{hyd} + \boldsymbol{F}_{p,i}^{ext}, \\
I_{p,i} \frac{\text{d} \boldsymbol{\omega}_{p,i}}{\text{d} t} &= \boldsymbol{T}_{p,i} = \boldsymbol{T}_{p,i}^{col} + \boldsymbol{T}_{p,i}^{hyd}.
\end{align}
Here, $m_{p,i} = \rho_{p,i} V_{p,i}$ is the mass of the particle with density $\rho_{p,i}$ and volume $V_{p,i}$, and $I_{p,i} = \tfrac{2}{5} m_{p,i} R_{p,i}^2$ is the moment of inertia for a sphere of radius $R_{p,i}$.
The temporal evolution of the particle's translational and angular velocity is thus given by the total force $\mathbf{F}_{p,i}$ and torque $\mathbf{T}_{p,i}$, respectively.
These contain contributions from inter-particle collisions, $\boldsymbol{F}_{p,i}^{col}$ and $\boldsymbol{T}_{p,i}^{col}$, and from hydrodynamic interactions, $ \boldsymbol{F}_{p,i}^{hyd}$ and $\boldsymbol{T}_{p,i}^{hyd}$, see Eqs.~\eqref{eq:hydrodynamic_force} and \eqref{eq:hydrodynamic_torque}.
Additionally, external forces $\boldsymbol{F}_{p,i}^{ext}$ might be present like gravity and buoyancy, given as $V_{p,i}(\rho_{p,i}-\rho_f) \boldsymbol{g}$, with the gravitational acceleration $\boldsymbol{g}$.

The time integration of these equations with a time step size $\Delta t_p$ is here accomplished with a Velocity Verlet scheme, which yields the explicit update formulas for the particle's position and velocity as~\cite{wachs2019}
\begin{align}
\boldsymbol{x}_{p,i}(t+\Delta t_p) &= \boldsymbol{x}_{p,i}(t) + \Delta t_p \boldsymbol{u}_{p,i}(t) + \tfrac{\Delta t_p^2}{2m_{p,i}} \boldsymbol{F}_{p,i}(t), \label{eq:velocity_verlet_1} \\
\boldsymbol{u}_{p,i}(t+\Delta t_p) &= \boldsymbol{u}_{p,i}(t)  + \tfrac{\Delta t_p}{2m_{p,i}} \left(\boldsymbol{F}_{p,i}(t)+\boldsymbol{F}_{p,i}(t+\Delta t_p)\right), \label{eq:velocity_verlet_2}
\end{align}
where $\boldsymbol{F}_{p,i}(t+\Delta t_p)$ is computed with the already updated position.
Analogously, the angular velocity and, if needed, the rotation are updated.

\subsubsection{Contact model for inter-particle collisions}

The discrete element method (DEM) is a soft-contact model that splits the inter-particle collisions into a normal and tangential part for each pairwise interaction:
\begin{align}
\boldsymbol{F}_{p,i}^{col} &= \sum_{j,j \neq i} \left( \boldsymbol{F}_{ij,n}^{col} + \boldsymbol{F}_{ij,t}^{col} \right) \label{eq:collision_force_total}\\
\boldsymbol{T}_{p,i}^{col} &= \sum_{j,j \neq i} (\boldsymbol{x}_{ij}^{cp}-\boldsymbol{x}_{p,i})\times \boldsymbol{F}_{ij,t}^{col} \label{eq:collision_torque_total}
\end{align}
Here, the normal part of the collision force acting on particle $i$ is given by a linear spring-dashpot model~\cite{cundall1979} as
\begin{equation}
\boldsymbol{F}_{ij,n}^{col} = -k_n \delta_{ij,n} \boldsymbol{n}_{ij} - d_n \boldsymbol{u}_{ij,n}^{cp}, \label{eq:normalCollisionForce}
\end{equation}
where $k_n$ and $d_n$ are the normal stiffness and damping coefficients, respectively.
Instead of attempting to specify those directly
, it is more convenient to introduce the collision time $T_c$ and the coefficient of restitution $e_\text{dry}$ \cite{vanDerHoef2006}.
The latter is defined as 
\begin{equation}
e_\text{dry} = -\frac{(\boldsymbol{u}_{ij,n} \cdot \boldsymbol{n}_{ij} )|_\text{post}}{(\boldsymbol{u}_{ij,n} \cdot \boldsymbol{n}_{ij} )|_\text{pre}}
\end{equation}
and is thus the ratio of the normal velocity after and before a single collision.
The collision time $T_c$ denotes the duration of a single collision event in dry conditions, i.e. in the absence of a fluid.
Since we are using a linear spring-dashpot model, the motion of the linear harmonic oscillator can be computed analytically by requiring that there is no overlap at the end of the collision. 
This yields the following relations \cite{vanDerHoef2006}:
\begin{equation}
k_n = \frac{m_{ij,\text{eff}}(\pi^2+\ln^2e_\text{dry})}{T_c^2},\quad d_n = -\frac{2{m_{ij,\text{eff}}\ln e_\text{dry}}}{T_c},
\label{eq:normalDEMparameters}
\end{equation}
with the effective mass given as
\begin{equation}
m_{ij,\text{eff}} = 
\begin{cases} 
\frac{m_{p,i}m_{p,j}}{m_{p,i}+m_{p,j}},& \quad \text{sphere-sphere,} \\
m_{p,i},& \quad \text{sphere-wall.}
\end{cases}
\end{equation}

As we will see in Sec.~\ref{sec:sphereWallCollision}, the collision time can significantly influence the collision behavior for immersed particles and has to be chosen with care.
In real collisions, the duration of a collision is several orders of magnitudes smaller than the typical time scales of the fluid.
The collision time in our model has to be seen as a numerical parameter, used to stretch the collision in time to allow for a proper treatment of the collision and the flow field without requiring extremely small time step sizes.
A longer collision time allows the surrounding fluid to better adapt to the sudden change in the particle's velocity but, at the same time, increases the surface overlap, which, however, should be kept small~\cite{costa2015,biegert2017}.

Similarly, the tangential collision force is given by a linear spring-dashpot model~\cite{thornton2013} as
\begin{equation}
\boldsymbol{F}_{ij,t}^{col,SD} =  - k_t \boldsymbol{\delta}_{ij,t} - d_t \boldsymbol{u}_{ij,t}^{cp}, \label{eq:tangentialCollisionForceSD}
\end{equation}
where $k_t$ and $d_t$ are the tangential stiffness and damping coefficients, respectively. The tangential spring displacement $\boldsymbol{\delta}_{ij,t}$ denotes the relative tangential motion between two particles that is accumulated starting from the time step of the impact $t_i$:
\begin{equation}
\boldsymbol{\delta}_{ij,t} = \int_{t_i}^{t} \boldsymbol{u}_{ij,t}^{cp}(t') \text{d}t' \label{eq:tangentialDisplacement}
\end{equation}
The actual tangential collision force is limited by the Coulomb friction model and thus given as
\begin{equation}
\boldsymbol{F}_{ij,t}^{col} = \min(\|\boldsymbol{F}_{ij,t}^{col,SD}\|, \| \mu_p \boldsymbol{F}_{ij,n}^{col} \|) \boldsymbol{t}_{ij}, \label{eq:tangentialCollisionForce}
\end{equation}
where $\mu_p$ is the (dynamic) coefficient of friction and 
\begin{equation}
	\boldsymbol{t}_{ij} = \boldsymbol{F}_{ij,t}^{col,SD}/ \|\boldsymbol{F}_{ij,t}^{col,SD}\|
\end{equation}
is the tangential unit vector.

In each time step of the discretized time integration of Eq.~\eqref{eq:tangentialDisplacement}, the former tangential displacement has to be projected into the current collision plane via
\begin{equation}
\tilde{\boldsymbol{\delta}}_{ij,t} = \boldsymbol{\delta}_{ij,t}(t) - \left( \boldsymbol{\delta}_{ij,t}(t) \cdot \boldsymbol{n}_{ij} \right)\boldsymbol{n}_{ij}
\end{equation}
and then rescaled to maintain the length.
Additionally, the displacement should not increase once the two surfaces are slipping and thus has to be reset to be conform with the Coulomb friction force~\cite{luding2008,biegert2017}.
This results in the update procedure
\begin{equation}
\boldsymbol{\delta}_{ij,t}(t+\Delta t_p) = 
\begin{cases}
\frac{\|\boldsymbol{\delta}_{ij,t}(t)\|}{\|\tilde{\boldsymbol{\delta}}_{ij,t}\|}\tilde{\boldsymbol{\delta}}_{ij,t} + \Delta t_p \boldsymbol{u}_{ij,t}^{cp},& \quad\text{if } \|\boldsymbol{F}_{ij,t}^{col,SD}\| < \| \mu_p \boldsymbol{F}_{ij,n}^{col} \| \\
-\frac{1}{k_t} \left(\|\mu_p \boldsymbol{F}_{ij,n}^{col}\|\boldsymbol{t}_{ij} + d_t \boldsymbol{u}_{ij,t}^{cp}\right),& \quad\text{else.} \\
\end{cases}
\end{equation}
For a new collision pair, the tangential displacement is initialized to zero~\cite{luding2008}.

The tangential stiffness and damping coefficients are related to the normal ones via \cite{thornton2013}
\begin{equation}
k_t = \kappa_p k_n,\quad d_t = \sqrt{\kappa_p}d_n, \label{eq:tangentialCollisionParameters} 
\end{equation}
which introduces the parameter
\begin{equation}
\kappa_p = \frac{2(1-\nu_p)}{2-\nu_p},
\end{equation}
with $\nu_p$ being the Poisson's ratio, a material property.

The advantage of this parameterization of the tangential collision model is that it does not require additional model parameters like a tangential coefficient of restitution~\cite{costa2015} or a critical impact angle~\cite{kempe2012} that have to be determined via specific laboratory experiments beforehand.
Instead, only the well-studied material property Poisson's ratio has to be specified.
Also, our parameterization is different from \cite{vanDerHoef2006,costa2015}, who made the assumption of a tangential collision time and equated this to the normal collision time, necessitating the definition of an effective tangential mass.

If required by the physical system, a straightforward extension of this contact model to also account for static friction is possible as shown in Ref.~\citenum{biegert2017}.
Also, adhesive interactions could be added in a modular fashion~\cite{benseghier2020}.

\subsection{Complete four-way coupled fluid-particle interaction algorithm}
\label{sec:complete_algorithm}

\begin{algorithm}[t]
	\begin{algorithmic}
	\FOR{each time step $t$}
	\STATE Apply fluid boundary conditions, Eq.~\eqref{eq:MEM_CLI}.
	\STATE Compute fluid-particle interactions, Eqs.~\eqref{eq:interaction_force} and \eqref{eq:interaction_torque}, and average over two time steps.
	\STATE Perform LBM step, Eqs.~\eqref{eq:LBM_Collide} and \eqref{eq:LBM_Stream} with MRT collision operator Eq.~\eqref{eq:MRT}.
	\FOR{each DEM subcycle}
	\STATE{Update particle position and rotation, Eq.~\eqref{eq:velocity_verlet_1}.}
	\STATE{Evaluate lubrication corrections, Eqs.~\eqref{eq:lubrication_correction_force_total} and \eqref{eq:lubrication_correction_torque_total}.}
	\STATE{Evaluate collision forces, Eqs.~\eqref{eq:collision_force_total} and \eqref{eq:collision_torque_total}.}
	\STATE{Set external forces, like gravity and buoyancy.}
	\STATE{Update particle translational and rotational velocity, Eq.~\eqref{eq:velocity_verlet_2}.}
	\ENDFOR
	\STATE Update the particle mapping into the fluid domain and reconstruct PDF information if necessary, Eq.~\eqref{eq:PDF_reconstruction}.
	\ENDFOR
	\end{algorithmic}
	\caption{Simulation approach for fluid-particle systems.}
	\label{alg:complete_algorithm}
\end{algorithm}

The complete algorithm that combines the previously presented building blocks, see Secs.~\ref{sec:lbm}, \ref{sec:coupling}, \ref{sec:lubrication_model}, and \ref{sec:rpd}, is shown in Alg.~\ref{alg:complete_algorithm}.
It features a subcycling loop which allows having a finer temporal resolution of the particle simulation, including the evaluation of collision and lubrication correction forces, without decreasing the overall time step size.
Due to this subcycling, the time step size for the particle integration is $\Delta t_p = \Delta t / n_\text{sub}$ where $n_\text{sub}$ is the number of subcycles.
This accounts for the fact that the usual time scales of particle interaction are smaller than of the fluid and is a common approach for coupled simulations \cite{costa2015,biegert2017,kidanemariam2014interface}.
During the subcycles, the fluid-particle interactions $\boldsymbol{F}_{p,i}^{fp}$ and $\boldsymbol{T}_{p,i}^{fp}$ remain constant.

Furthermore, we apply an averaging of these two quantities over the current and previous time step to even out fluctuations that are inherent to LBM due to its bounce-back nature of the boundary conditions ~\cite{ladd1994_2}.

\subsection{Sphere freely settling under gravity}

\label{sec:tenCateSettling}

\subsubsection{Background}

This test is based on the experiments and simulations performed by Ten Cate \textit{et al.}~\cite{tenCate2002} and features a single sphere settling freely inside a box filled with different silicon oils.
Due to its simplicity, it has been adopted by others \cite{seil2018,biegert2017} and has thus become one of the standard tests for moving particle simulations.
It is well suited to validate the fluid-particle interaction and the time integrator for the particle motion as it features the acceleration and deceleration phases of the sphere at the beginning and when close to the bottom wall, respectively.
The temporal evolution of the distance from the bottom wall, as well as the settling velocity, is evaluated and compared against the experimental results \cite{tenCate2002}.

\subsubsection{Description}

The simulation features a box of size $0.1$\,m$\times0.1$\,m$\times0.16$\,m with a sphere of diameter $D_p = 0.015$\,m and density $\rho_p = 1120$\,kg/m$^3$ and a gravitational acceleration of $g = 9.81$\,m/s$^2$.
The properties of the four different fluids are given in Tab.~\ref{tab:tenCate}, together with the experimentally measured maximum settling velocity $u_s$ and the Reynolds number.
In the simulation, a domain size of $135\times135\times216$ cells is chosen, resulting in $D_p/\Delta x = 20.25$.
Additionally, we use $u_s = 0.02$ in lattice units and apply the density ratio from the experiments, which then determines all other simulation quantities accordingly.
The initial distance of the sphere surface from the bottom wall is $8.2 D_p$ and the simulation is stopped right before the sphere hits the bottom wall.

\begin{table}[t]
	\caption{Fluid parameters in physical units and the resulting $Re$ for the single sphere settling test, used in the experiments by Ref.~\citenum{tenCate2002}.}
	\label{tab:tenCate}
	\centering
	\begin{tabular}{l||ccc|c}
		Case & $\rho_f$ / (kg/m$^3$) & $\mu_f$ / (Ns/m$^2$) & $u_s$ / (m/s)  & $Re$ \\\hline
		TC1 & 970 & 0.373 & 0.036 & 1.4 \\
		TC2 & 965 & 0.212 & 0.057 & 3.9 \\
		TC3 & 962 & 0.113 & 0.087 & 11.1 \\
		TC4 & 960 & 0.058 & 0.122 & 30.3 \\
	\end{tabular}
\end{table}

\subsubsection{Results and Discussion}

\begin{figure}[t]
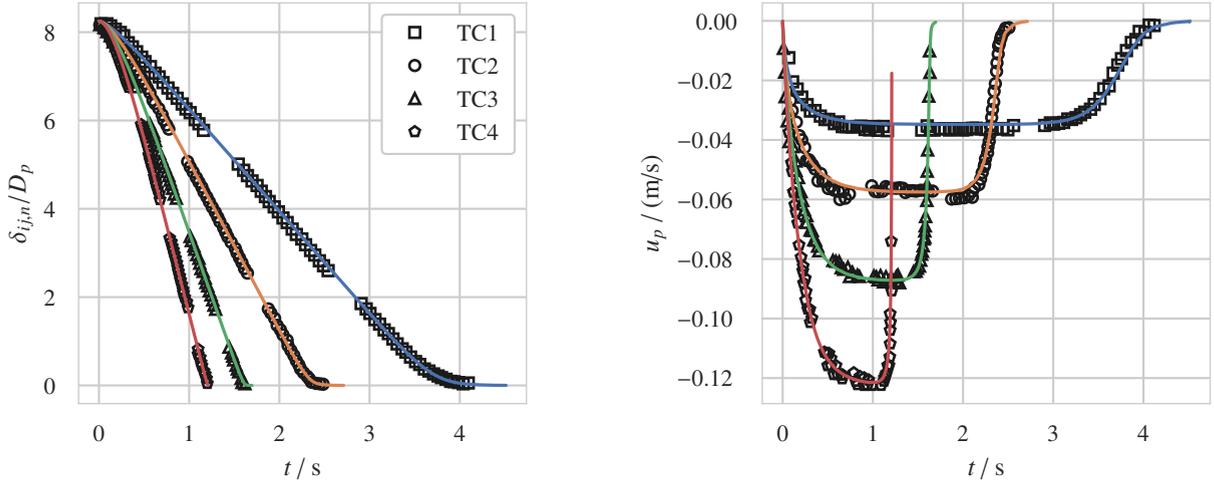

	\centering
	\begin{subfigure}[b]{0.5\textwidth}
		\input{figures/settling_sphere_height_test.pgf}
	\end{subfigure}~
	\begin{subfigure}[b]{0.5\textwidth}
		\input{figures/settling_sphere_vel_test.pgf}
	\end{subfigure}
	\caption{Comparison of the settling behavior at four different Reynolds numbers of a sphere between the experimental data from \cite{tenCate2002} to our simulations, given by the colored lines. The dimensionless gap width (left) and the settling velocity (right) are shown over time.}
	\label{fig:tenCate_settling_test}
\end{figure}


A comparison between the simulation and the experimental results is given in Fig.~\ref{fig:tenCate_settling_test}, showing the gap width and the particle velocity in settling direction, and an excellent agreement can be seen for all four cases.
In particular, the initial acceleration and final deceleration in the proximity of the wall are well captured.
From this, we can conclude that our coupling approach is able to predict the settling behavior at moderate Reynolds numbers accurately.

The test case of Ref.~\citenum{uhlmann2014} could be used for validating settling at higher Reynolds numbers.
For a very similar coupling approach to the one applied here, this has already been used and demonstrated in Ref.~\citenum{rettinger2017}, and is thus not repeated here.
However, we explicitly note here that these settling tests do not contain particle-particle or particle-wall collisions.
Consequently, they cannot be taken as the sole validation scenario if collisions might occur during the simulation and influence the dynamics of the physical system.
For that purpose, the setup in the following section has to be applied. 

\subsection{Collision dynamics of a sphere impacting normally onto a wall}

\label{sec:sphereWallCollision}

\subsubsection{Background}

The models to account for particle-particle and particle-wall interactions typically come with several parameters for which suitable values have to be determined.
For simulations of submerged particles, those interactions are not only described by the collision resolution model, as here the DEM, for direct contact but also by the fluid-particle coupling algorithm and the lubrication correction model.
At best, those parameters are physical material parameters like the coefficient of restitution $e_\text{dry}$ that can be obtained from corresponding experimental measurements or material tables.
This is unfortunately usually not possible for all parameters and these have thus to be calibrated and validated.

For this purpose, the collision of a single sphere with a wall inside a fluid-filled domain is the simplest test case, resulting in a \textit{wet} collision.
In contrast to the case without fluid, called \textit{dry} collision, the apparent coefficient of restitution changes depending on the fluid and the sphere properties as well as the impact velocity $u_s$ \cite{gondret2002}.
This dependence is captured concisely by the Stokes number, given as
\begin{equation}
\textit{St} = \frac{1}{9}\frac{\rho_p}{\rho_f}\frac{u_s D_p }{\nu_f}.
\end{equation}
Several experimental studies exist that report the wet coefficient of restitution $e_\text{wet}$ as a function of the Stokes number \cite{joseph2001,gondret2002,yang2006} and it might thus be tempting to use those as reference.
However, determining the coefficient of restitution requires the evaluation of the pre- and post-collision velocity of the sphere.
Those values depend significantly on the exact time when the evaluation is done as the sphere's velocity will be affected and altered significantly by the always acting hydrodynamic interaction forces, in contrast to measurements for the dry case \cite{izard2014}.
This leads to a somewhat diffuse point cloud when combining the findings of all these experiments.
Additionally, the same issue arises for simulations which has also been noted in e.g. Ref.~\citenum{kidanemariam2014interface} who also demonstrated the considerable influence of the evaluation timing on the measured $e_\text{wet}$.
Such a procedure would thus introduce an undesired additional parameter, or degree of arbitrariness, which does not permit a rigorous calibration of our model parameters.

Instead, for the case of a wet sphere-wall collision in the normal direction, Gondret \textit{et al.}~\cite{gondret2002} also provide some experimentally measured rebound trajectories after the collision.
These form a well-suited reference for numerical tests that aim to reproduce those experiments.
Such comparative studies have already been carried out for classical DNS approaches with the immersed boundary method \cite{kempe2012, costa2015, biegert2017,jain2019} but we are not aware of any attempts using LBM.
We will thus use this test case to study the effect of the collision time $T_c$ used in the parameterization of the DEM, Eq.~\eqref{eq:normalDEMparameters}, in detail and to establish a guideline for choosing it for our approach.
Furthermore, the minimal admissible gap size for the lubrication force, $\delta_{n,\text{min}}^\text{lub}$ from Eq.~\eqref{eq:lubrication_correction_gap_size}, and the effect of subcycles $n_\text{sub}$, see Sec.~\ref{sec:complete_algorithm}, will be determined.
This will finalize the calibration of the interaction model in the normal direction which is then also validated at last.

\subsubsection{Description}

In this test case, the sphere is settling inside a box and ultimately hits the wall at the bottom of the domain.
This results in a rebound of which the trajectory can be compared against experimental data from Ref.~\citenum{gondret2002}.
As mentioned before, the setup is characterized by the Stokes number \textit{St}, which thus has to match with the corresponding experiments.
Initially, we place the sphere horizontally centered and at a distance of $\tfrac{3}{2}D_p$ from the top boundary.
Since not all spheres reached their terminal velocity and an exact reproduction of the experimental setup would be computationally permissive, we decided to follow an approach similar to Ref.~\citenum{biegert2017}.
Thus, we artificially accelerate the sphere to the desired velocity by imposing the velocity via Eq.~\eqref{eq:artificial_acceleration}, using $c_{acc}=10$ as in Sec.~\ref{sec:BulkViscEffect}.
When the sphere is close to the bottom wall, i.e., when the gap size is smaller than $D_p$, we turn off the artificial acceleration and let it settle freely under the action of gravity and buoyancy, with a gravitational acceleration of $g=9.81$m/s$^2$.
This is done to not perturb the hydrodynamic interaction of the sphere with the wall before the collision.
The physical parameters and the domain sizes for the different setups are stated in Tab.~\ref{tab:sphereWallCollision}.


\begin{table}[t]
	\caption{Parameters for the sphere-wall collision setup, used in the experiments by Ref.~\citenum{gondret2002}.}
	\label{tab:sphereWallCollision}
	\centering
	\begin{tabular}{c||ccccccc}
		\textit{St} & $Re$ & $\rho_f$ / (kg/m$^3$) & $\mu_f$ / (Ns/m$^2$) & $D_p$ / m & $\rho_p$  / (kg/m$^3$)  & domain size / $D_p$ & $e_\text{dry}$\\\hline
		27 & 30 & 965 & 0.1 & 0.006 & 7800 & $12\times12\times48$ & 0.97  \\
		100 & 110 & 953 & 0.02 & 0.004 & 7800 & $12\times12\times48$ & 0.97  \\
		152 & 164 & 935 & 0.01 & 0.003 & 7800 & $12\times12\times64$ & 0.97 \\
	\end{tabular}
\end{table}


\subsubsection{Results}

In the following parts, the effect of various parameters onto the collision behavior are studied.
For a better overview of the applied parameterization, the parameter sets are summarized in Tab.~\ref{tab:sphereWallCollision_parameters}.

\begin{table}[t]
	\caption{Numerical parameter sets for simulations of the sphere-wall collision test.}
	\label{tab:sphereWallCollision_parameters}
	\centering
	\begin{tabular}{l||ccccc}
		Case & $T_c$ & $D_p$ & $u_s$ & $n_\text{sub}$ & $\delta_{n,\text{min}}^\text{lub}$ \\\hline
		SW1 & $[1,...,200]$ & 20 & 0.02 & 50 & 0.001 $R_p$ \\
		SW2 & $[1,...,200]$ & 30 & 0.02 & 50 & 0.001 $R_p$ \\
		SW3 & $[1,...,200]$ & 40 & 0.02 & 50 & 0.001 $R_p$ \\
		SW4 & $[1,...,200]$ & 20 & 0.01 & 50 & 0.001 $R_p$ \\ \hline
		SW5 & Eq.~\eqref{eq:collisionTimeChoice} & 20 & 0.02 & 50 & $[0.001,...,0.0024]R_p$  \\
		SW6 & Eq.~\eqref{eq:collisionTimeChoice} & 10 & 0.02 & 50 & $[0.001,...,0.0024]R_p$ \\
		SW7 & Eq.~\eqref{eq:collisionTimeChoice} & 30 & 0.02 & 50 & $[0.001,...,0.0024]R_p$ \\
		SW8 & Eq.~\eqref{eq:collisionTimeChoice} & 40 & 0.02 & 50 & $[0.001,...,0.0024]R_p$ \\ \hline
		SW9 & Eq.~\eqref{eq:collisionTimeChoice} & 20 & 0.02 & $[1,...,50]$ & Eq.~\eqref{eq:lubricationGapSize}\\ \hline
		SW10 & Eq.~\eqref{eq:collisionTimeChoice} & 20 & 0.02 & 10 & Eq.~\eqref{eq:lubricationGapSize}
	\end{tabular}
\end{table}

\paragraph{Effect and choice of collision time}

\begin{figure}[t]
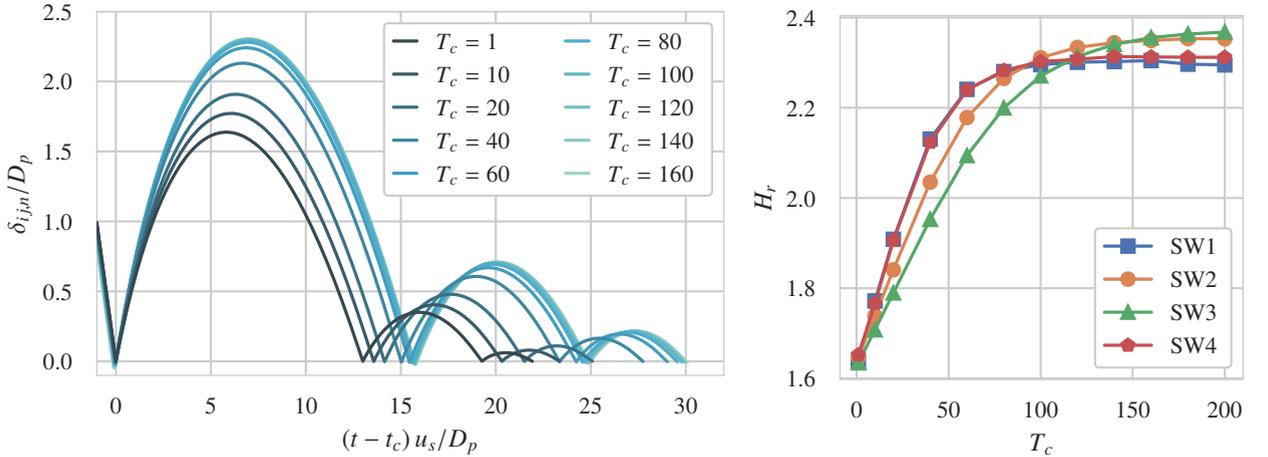

	\centering	
	\begin{subfigure}[t]{0.59\textwidth}
	\input{figures/sphereWallCollision_St152_Tc_D20_u002.pgf}
	\caption{Normalized gap size over normalized time for the parameter set SW1.}
	\label{fig:sphereWallCollision_St152_Tc_D20_u002}
	\end{subfigure}~
\hfill
\begin{subfigure}[t]{0.4\textwidth}
		\input{figures/sphereWallCollision_St152_reboundHeight_over_Tc.pgf}
	\caption{Rebound height over collision time for different parameter sets.}
	\label{fig:sphereWallCollision_St152_reboundHeight_over_Tc}
\end{subfigure}
\caption{Effect of the collision time $T_c$ on the sphere-wall collision with $\textit{St}=152$.}
	
\end{figure}

As discussed in Sec.~\ref{sec:rpd}, the collision time $T_c$ determines the duration of the contact-based collision.
To study the general effect of this parameter, we use $\textit{St}=152$ and vary $T_c$ for the parameter set SW1 from Tab.~\ref{tab:sphereWallCollision_parameters}.
The resulting trajectories, given as gap sizes between the lower sphere surface and the bottom plane, are shown in Fig.~\ref{fig:sphereWallCollision_St152_Tc_D20_u002}.
There, the time is shifted by $t_c$, i.e., the moment when there is no contact any more between sphere and wall after the first collision, and additionally normalized.
It can be seen that the collision behavior exhibits a strong dependence on the collision time with larger values for $T_c$ resulting in higher rebounds.
The difference between the trajectories for the large collision times get smaller to the point where the curves for the highest tested collision times collapse to a single one.
For a more detailed evaluation, we extract the rebound height $H_r$ which is the largest value for the gap size after the first collision, i.e.
\begin{equation}
H_r = \max_t \delta_{ij,n}(t)/D_p, \text{ for } t > t_c,
\label{eq:rebound_height}
\end{equation}
and show the results in Fig.~\ref{fig:sphereWallCollision_St152_reboundHeight_over_Tc}.
For the parameter set SW1, we observe that $H_r$ reaches its highest and final value from around $T_c=80$ onward.
Following the general guidelines regarding $T_c$, stated in Sec.~\ref{sec:rpd}, it can be concluded that for this case the fluid requires at least 80 time steps to accommodate the sudden change in the particle velocity and to avoid an otherwise strong damping of the collision via hydrodynamic interaction.
To also avoid a too long artificial stretching of the collision in time, which would then also increase the penetration depth of the surfaces during the collision, larger values of $T_c$ are undesired and we thus propose to use $T_c=80$ for this case. 

To investigate whether this value is affected by the choice of the overall spatial or temporal resolution, we repeat this study with parameter sets SW2, SW3 and SW4 where the diameter and the settling velocity in lattice units are changed, respectively.
Consequently, the spatial resolution is increased for SW2 and SW3, whereas the temporal resolution is increased for SW4.
The outcome is also included in Fig.~\ref{fig:sphereWallCollision_St152_reboundHeight_over_Tc}.
For SW2 and SW3, we observe a qualitatively similar behavior as for SW1, but a higher value of $T_c$ is required to reach a saturated value of $H_r$.
In particular, approximately $T_c=120$ is needed for SW2 and $T_c=160$ for SW3. 
This suggests that for finer resolutions, the collision time should be adapted as well to higher values, i.e., $T_c = f(D_p)$ with a function $f$ that is linear in $D_p$.
We note that the slight variations of the saturated values of $H_r$ in these cases originate from the influence of the lubrication gap size parameter, as explained in the next part.

On the other hand, the case SW4, with the smaller physical time step size determined by a smaller $u_s$ in lattice units, exhibits almost identical trajectories in comparison to SW1 while at the same time reducing the maximum penetration depth.
This shows that the choice of the collision time should be independent of $u_s$.
The collision time is thus decoupled from the physical time scale and should be seen as a numerical parameter.
Instead, we here suggest to make use of the lattice speed of sound $c_s = 1/\sqrt{3}$, which is another characteristic quantity in LBM simulations, see Sec.~\ref{sec:lbm}.
This quantity is related to the speed at which information is transported inside the fluid.
It, therefore, fits the observation that a certain number of time steps are required such that the flow field around the colliding sphere is aware of the sudden change in the velocity and can adapt to it.

From these observations, we obtain the following guideline for choosing the collision time:
\begin{equation}
T_c = 2.31 \frac{D_p}{c_s}.
\label{eq:collisionTimeChoice}
\end{equation} 





\paragraph{Minimal lubrication gap size}


\begin{figure}[t]
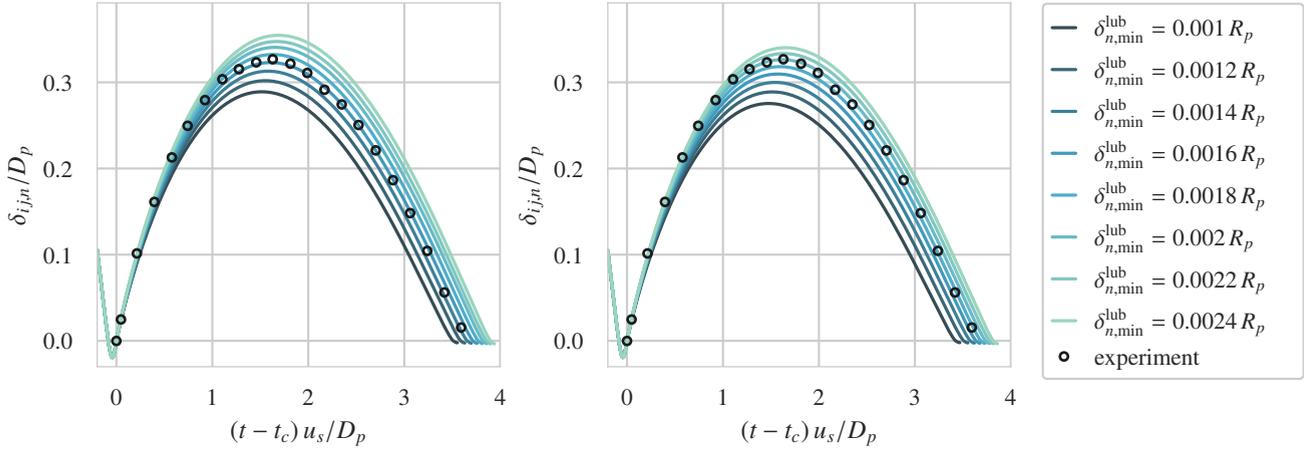

	\centering
	\begin{subfigure}[b]{0.4\textwidth}
		\input{figures/sphereWallCollision_St27_gs_D20.pgf}
	\end{subfigure}~
	\begin{subfigure}[b]{0.59\textwidth}
		\input{figures/sphereWallCollision_St27_gs_D30.pgf}
	\end{subfigure}
	\caption{Normalized gap size over normalized time of the sphere-wall collision with $\textit{St}=27$ for varying minimal admissible gap sizes $\delta_{n,\text{min}}^\text{lub}$. The parameter sets are SW5 (left) and SW7 (right). The reference experimental data is taken from Ref.~\citenum{gondret2002}.}
	\label{fig:sphereWallCollision_St27_gs}
\end{figure}

As a next step, we will determine the parameter for the minimal lubrication gap size, $\delta_{n,\text{min}}^\text{lub}$ from Eq.~\eqref{eq:lubrication_correction_gap_size}, which is part of the lubrication correction.
For that, we use $\textit{St}=27$ since for that case lubrication forces play a more dominant role in the collision behavior.
The influence of this parameter on the trajectory can be seen in Fig.~\ref{fig:sphereWallCollision_St27_gs} on the left.
If smaller gap sizes are allowed before they are capped, the lubrication forces become larger.
As they are opposing the motion, the collision behavior is damped and the rebound is lower.
Larger values then result in higher rebound trajectories.
By comparison with the experimental reference data, we find that  $\delta_{n,\text{min}}^\text{lub}=0.0017\,R_p$ provides the best agreement here.

To investigate the behavior for varying diameters, we also carried out simulations with $D_p=10, 30$, and $40$, for which $D_p=30$ is shown in Fig.~\ref{fig:sphereWallCollision_St27_gs} on the right.
The same overall behavior can be seen and, in comparison to $D_p=20$, the trajectories are shifted downwards.
Consequently, larger values for $\delta_{n,\text{min}}^\text{lub}$ are required to achieve the match with the experiments, here given by $\delta_{n,\text{min}}^\text{lub}=0.002\,R_p$ .

\begin{table}[t]
	\caption{Best fit value for minimal lubrication gap size for different diameters. Additionally, the values from the proposed relation are given.\label{tab:sphereWallCollision_St27_gs_bestFit}}
	\centering
	\begin{tabular}{cc|c|c}
		case & $D_p$ & best fit $\delta_{n,\text{min}}^\text{lub}$ & Eq.~\eqref{eq:lubricationGapSize}\\\hline
		SW6 & 10 & 0.007 & 0.007\\ 
		SW5 & 20 & 0.017 & 0.017\\
		SW7 & 30 & 0.030  & 0.031\\
		SW8 & 40 & 0.048 & 0.048\\
	\end{tabular}
\end{table}

The same results can be drawn from $D_p=10$ and $40$ as well and we summarized the corresponding values that lead to the best agreement in Tab.~\ref{tab:sphereWallCollision_St27_gs_bestFit} for all four cases.
Based on this table, we refine the scaling with the radius and propose to use the relation
\begin{equation}
\delta_{n,\text{min}}^\text{lub}=(0.001 + 0.00007R_p)\,R_p
\label{eq:lubricationGapSize}
\end{equation} 
that recovers the found values for the best fit very well, see Tab.~\ref{tab:sphereWallCollision_St27_gs_bestFit}.



\paragraph{Sensitivity to number of subcycles}

In this last calibration test, we investigate the sensitivity of the rebound trajectory with respect to the number of subcycles taken by the rigid body simulation, i.e., $n_\text{sub}$.
For this study, we use $\textit{St}=27$ with the parameter set SW9 in Tab.~\ref{tab:sphereWallCollision_parameters}.
We randomly perturb the initial sphere position by one cell size in $z$-direction and carry out ten different runs.
For each realization, we then evaluate the rebound height $H_r$ from Eq.~\eqref{eq:rebound_height}.
We note that in all cases, the same Stokes number is used and thus the same collision with the same rebound trajectories should be observed.
Possible differences in the results thus primarily originate from the sampling rate for lubrication correction as well as for collision detection and resolution.
The result of this study is given in Tab.~\ref{tab:sphereWallCollision_nsub_D20}.
It can be seen that simulations without substepping, i.e. $n_\text{sub}=1$, suffer from a relatively large variance in the rebound height and thus in the overall rebound behavior.
This variance can be decreased effectively with subcycling, where only around $5$ sub steps are needed to obtain differences below $1\%$.
Furthermore, subcycling increases the overall accuracy of the collision treatment which can be observed in the average rebound height that is essentially the same once $10$ or more sub steps are applied.
We thus propose to use at least $n_\text{sub} = 10$ for this setup.


\begin{table}[t]
	\centering
	\caption{Sensitivity of the rebound height to the number of subcycles $n_\text{sub}$. Minimum, maximum and average rebound heights of $10$ distinct runs are given where the initial sphere position is randomly perturbed. The relative difference is calculated as the difference of the maximum and minimum rebound height, divided by the average one.}
	\begin{tabular}{c|ccc|c}
		$n_\text{sub}$ & $\min H_r$ & $\max H_r$ & average $H_r$ & rel. diff. in \% \\\hline
	1 & 0.3135 & 0.3483 & 0.3218 &  10.8\\
	2 & 0.3177 & 0.3350 & 0.3263 &   5.3\\
	5 & 0.3264 & 0.3270 & 0.3267 &   0.2\\
	10 & 0.3269 & 0.3276 & 0.3274 &   0.2\\
	20 & 0.3274 & 0.3280 & 0.3277 &   0.2\\
	30 & 0.3265 & 0.3281 & 0.3276 &   0.5\\
	40 & 0.3274 & 0.3279 & 0.3277 &   0.2\\
	50 & 0.3270 & 0.3281 & 0.3277 &   0.3\\
	\end{tabular}
	\label{tab:sphereWallCollision_nsub_D20}
\end{table}

\paragraph{Validation}

\begin{figure}[t]
	\centering	
	\input{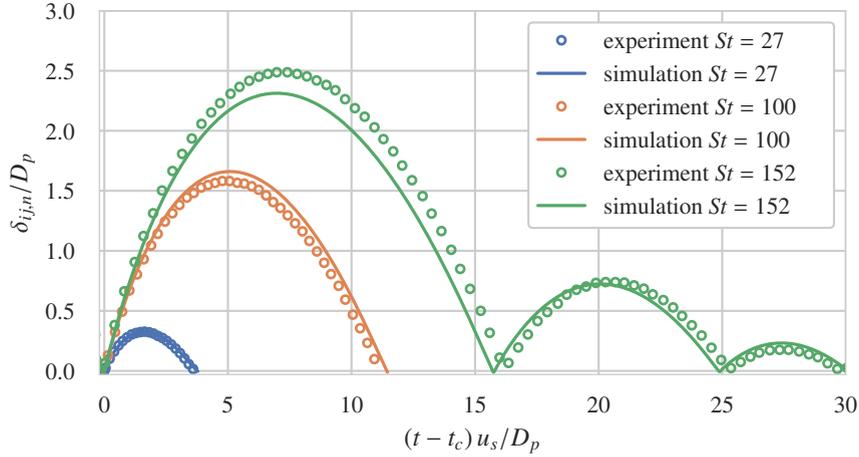}
	\caption{Validation study for different cases from Tab.~\ref{tab:sphereWallCollision}. The simulations use the parameter set SW10. The reference experimental data is taken from Ref.~\citenum{gondret2002}.}
	\label{fig:sphereWallCollision_validation}
\end{figure}

With the fully calibrated simulation at hand, we can finally conduct a validation test of our approach with the parameter set SW10 from Tab.~\ref{tab:sphereWallCollision_parameters}, which makes use of all found relations.
For that, we take the experimental results of $\textit{St}=100$ and $\textit{St}=152$ as a reference since those have not yet been used for the calibration.
The results, where we have also added the case $\textit{St}=27$ for completeness, are shown in Fig.~\ref{fig:sphereWallCollision_validation}.
From that, we see that the experimental and simulation trajectories agree well in those cases, with a small deviation for $\textit{St}=152$.

\subsubsection{Discussion}

We have seen that the collision time $T_c$ should be chosen large enough to avoid numerical damping effects by the fluid but as small as possible to avoid stretching the collision duration unnecessarily, which results in large penetration depths.
Based on the observable convergence-like behavior of the trajectories, we proposed to choose $T_c$ according to Eq.~\eqref{eq:collisionTimeChoice}, i.e., the smallest value for which the saturated rebound heights can be observed.
It should be noted that this argument is different from the one made in Ref.~\citenum{kempe2012}, where also the influence of $T_c$ on the collision behavior for this test setup has been observed.
There, however, $T_c=10\Delta t$ has been proposed as the value that best reproduces the trajectories of two test cases, i.e., as a fitting parameter.
The same choice is made in Ref.~\citenum{biegert2017}, whereas $T_c=8\Delta t$ is applied in Ref.~\citenum{costa2015}.
Generally, these values are considerably smaller than the ones used in the present work.
This can be explained by the fact that in LBM, the physical time step sizes are usually smaller than the ones applied in classical DNS approaches.
Consequently, the physical collision time is actually similar in both approaches, as well as the observed penetration depths.

The parameter for the minimal admissible gap size, $\delta_{n,\text{min}}^\text{lub}$, is here considered a fitting parameter, as in Ref.~\citenum{biegert2017}, and is a quadratic function of the sphere radius.
The dependence on the radius is often motivated by interpreting this parameter as a model for the presence of surface asperities \cite{izard2014,costa2015}. 
Then a linear dependence on the radius is used.
Having determined the minimum gap size and the cut-off distance for the lubrication correction in the normal direction, we can give a rough estimate of the numerical resolution that would be required to resolve the lubrication forces fully and without a model.
This is done by equating Eq.~\eqref{eq:lubricationGapSize} and Eq.~\eqref{eq:lubricationCutOff}, yielding $R_p\approx90$.
For most practical cases, such a fine resolution is computationally prohibitive, and thus, the lubrication correction model is a reasonable and efficient substitute.
We also note that without such a correction model and by using resolutions similar to the ones applied in our studies, the trajectories for low Stokes number collisions can not be captured correctly~\cite{costa2015,biegert2017}, which underlines their importance.

The here applied number of subcycles, $n_\text{sub} = 10$, is below usually reported ones of $50$ in Ref.~\citenum{costa2015} and $15$ in Ref.~\citenum{biegert2017}, which again can be addressed to the smaller fluid time step sizes in LBM.
We note that the only argument against using an unnecessarily large amount of subcycles is the overall computational performance of the algorithm.
Each sub step comes with additional costs for the lubrication force evaluation and collision detection, as well as for synchronization when executed in a parallel environment.
It is thus desired to choose $n_\text{sub}$ as small as possible to decreases the computing time.

Finally, we want to highlight a crucial difference between our collision model in the normal direction in comparison to existing ones.
It is a common approach to disregard the hydrodynamic interaction force during the collision \cite{kempe2012,costa2015,biegert2017} to reduce the drag experienced by the sphere.
This is motivated by the observation that with this approach, the resulting rebound trajectory is higher and thus the simulation can better capture the large Stokes number cases.
This can be seen in our validation study, Fig.~\ref{fig:sphereWallCollision_validation}, with $\textit{St}=152$ where we slightly underpredict the experimental trajectory.
At the same time, however, the trajectory for $\textit{St}=100$ would become higher as well, reducing the agreement between experiments and simulation.
Furthermore, special treatment of enduring contacts is required to avoid artifacts during the simulation of erosion processes with initially resting particles, which ultimately renders the resulting collision algorithm rather complex \cite{vowinckel2016,biegert2017}. 
Consequently, we decided not to introduce this procedure in our model.

\subsection{Oblique collision of sphere with wall}

\subsubsection{Background}

In this last test, the tangential part of the collision model will be validated for both, dry and wet, scenarios of oblique sphere-wall collisions.
Since, unfortunately, no experimental studies with such detailed trajectory information as for the normal collision from Sec.~\ref{sec:sphereWallCollision} are available, the standard approach is to compare the rebound angle as a function of the impact angle.

At first, the dry case, i.e., in the absence of a surrounding fluid, is used to generally establish the validity of the tangential collision model, Eq.~\eqref{eq:tangentialCollisionForce}, and the corresponding parameterization.
As mentioned in Sec.~\ref{sec:rpd}, the here used tangential collision model has the advantage that it requires a minimal set of material parameters, in contrast to others that, e.g., require a tangential coefficient of restitution~\cite{costa2015} or a critical impact angle~\cite{kempe2012}.
The experimental data from Ref.~\citenum{foerster1994} is used as a reference with the parameters given in Tab.~\ref{tab:obliqueCollision} corresponding to a glass sphere impacting on an aluminum plate.

In a second step, the fluid is included and the experimental findings from Ref.~\citenum{joseph2004} act as a reference.
It features a glass (TW1) and a steel (TW2) sphere dropped in different water-glycerol mixtures.

\begin{table}
	\caption{Physical parameter settings for the dry \cite{foerster1994} and wet \cite{joseph2004} oblique collision tests.}
	\centering
	\begin{tabular}{c||ccccc|cc}
		Case & $D_p$ / m & $\rho_p$ / (kg/m$^3$) & $e_\text{dry}$ & $\nu_p$ & $\mu_p$ & $\rho_f$ / (kg/m$^3$) & $\mu_f$ / (Ns/m$^2$) \\\hline
		TD & 0.00318 & 2500 & 0.83 & 0.22 & 0.12 & - & -  \\\hline
		TW1 & 0.00127 & 2540 & 0.97 & 0.23 & 0.15 & 1081 & 0.0025 \\ 
		TW2 & 0.00127 & 7780 & 0.97 & 0.27 & 0.02 & 1091 & 0.003  
	\end{tabular}
	\label{tab:obliqueCollision}
\end{table}

\subsubsection{Description}

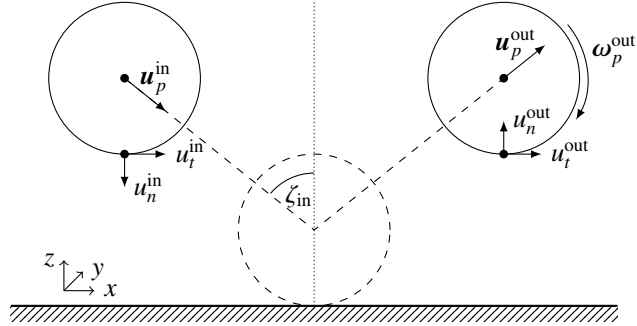
\begin{figure}
	\centering
	\begin{tikzpicture}
	\coordinate[] (xpi) at (-2.5,3);
	\coordinate[] (xpj) at (2.5,3);
	\coordinate[] (col) at (0,1);
	\coordinate[] (coordSystem) at (-3.3,0.2);
	
	\draw[->] (coordSystem) -- ++(0.4,0) node[pos=1,right]{$x$};
	\draw[->] (coordSystem) -- ++(0,0.4) node[pos=1,left]{$z$};
	\draw[->] (coordSystem) -- ++(0.25,0.25) node[pos=1,right]{$y$};
	
	\draw (xpi) circle (1);
	\draw[fill=black] (xpi) circle (0.05);
	
	\draw (xpj) circle (1);
	\draw[fill=black] (xpj) circle (0.05);
	
	\draw[dashed] (col) circle (1);
	
	\draw[thick] (-4,0) -- (4,0);
	\fill[pattern=north east lines] (-4,-0.2) rectangle (4,0);
	
	\draw[-latex] (xpi) -- ($(xpi)!0.7cm!(col)$) node[pos=0,right]{\ $\boldsymbol{u}_{p}^\text{in}$};
	\draw[-latex] (xpj) -- ($(xpj)!-0.7cm!(col)$) node[pos=1,left]{\ $\boldsymbol{u}_{p}^\text{out}$};
	\draw[-latex] ($(xpj) + (40:1.1)$) arc (40:-30:1.1) node[pos=0.3,right]{$\boldsymbol{\omega}_{p}^\text{out}$};
	
	\draw[densely dotted] (0,0) -- ++(0,4);
	\draw[dashed] (xpi) -- (col) --(xpj);
	
	\draw[fill=black] ($(xpi) + (-90:1)$) circle (0.05);
	\draw[-latex] ($(xpi) + (-90:1)$) -- ++(0.55,0) node[pos=1,right]{$u_{t}^\text{in}$};
	\draw[-latex] ($(xpi) + (-90:1)$) -- ++(0,-0.44) node[pos=1,right]{$u_{n}^\text{in}$};
	\draw[fill=black] ($(xpj) + (-90:1)$) circle (0.05);
	\draw[-latex] ($(xpj) + (-90:1)$) -- ++(0.5,0) node[pos=1,right]{$u_{t}^\text{out}$};
	\draw[-latex] ($(xpj) + (-90:1)$) -- ++(0,0.44) node[pos=1,right]{$u_{n}^\text{out}$};
	
	\draw($(col) + (140:0.75)$) arc (140:90:0.75);
	\node[] at (-0.2,1.45) {$\zeta_\text{in}$};
	\end{tikzpicture}
	\caption{Sketch for the oblique sphere-wall collision. The states \textit{in} and \textit{out} are taken right before and after the collision.}
	\label{fig:obliqueWallCollision_sketch}
\end{figure}

This test features a sphere that hits a plane at different angles $\zeta_\text{in}$ and the post-collision state is evaluated.
Before the collision, the sphere solely has a translational velocity $\boldsymbol{u}_{p}^\text{in}$.
Consequently, the tangential surface velocity of the point closest to the plane, i.e., the contact point, is $u_{t}^\text{in} = u_{p,t}^\text{in}$.
After the collision, a rotational velocity is present and the tangential surface velocity is given as $u_{t}^\text{out} = u_{p,t}^\text{out} - R_p \omega_{p,y}^\text{out}$.
We then define the following velocity ratios \cite{foerster1994}
\begin{equation}
\Psi_\text{in} = -\frac{u_{t}^\text{in}}{u_{n}^\text{in}}, \quad \Psi_\text{out} =-\frac{u_{t}^\text{out}}{u_{n}^\text{in}}, \label{eq:obliqueCollisionRatios}
\end{equation}
which results in $\Psi_\text{in} = \tan(\zeta_\text{in})$.
These value pairs can be compared to experimental results \cite{foerster1994,joseph2004}.
This setup and the relevant quantities are shown in Fig.~\ref{fig:obliqueWallCollision_sketch}.

For the dry case (TD), the simulation setup is straightforward and mimics the one given in Fig.~\ref{fig:obliqueWallCollision_sketch}.
Since no gravity is considered, the velocities before and after the collision remain constant and can be evaluated at any time.

In the wet cases (TW1 and TW2), it has to be noted that the experimental data in Ref.~\citenum{joseph2004} lack detailed information about the exact collision properties, e.g., given as a Stokes number.
Additionally, the experimental apparatus features a pendulum, which is challenging to reproduce in the simulations.
This also explains the different setups used in the literature \cite{kempe2012,biegert2017,jain2019} to reproduce this case adequately in simulations.
However, the common property of all these simulations is a high Stokes number, which is obtained by an artificially high gravitational acceleration.
 
To obtain more control over the simulation, we employ a different approach here.
We use a horizontally periodic box of size $[12\times12\times24]D_p$ and place the sphere at position $[6\times6\times22.5]D_p$. 
We then define and prescribe a Stokes number in the normal direction, 
\begin{equation}
	\textit{St}_n = \frac{1}{9}\frac{\rho_p}{\rho_f}\frac{|u_n^\text{in}| D_p }{\nu_f}.
\end{equation}
For that purpose and similar to before, we artificially accelerate the sphere by setting its velocity to 
\begin{align}
\boldsymbol{u}_p(t) & = \left(1 - \exp(-c_{acc}\,t/t_{\textit{St}})\right) (u_t^\text{in},0,u_n^\text{in})^\top, \\
\boldsymbol{\omega}_p(t) & = (0,0,0)^\top, \label{eq:artificial_oblique_acceleration}
\end{align}
with $c_{acc}=25$.
When the smallest distance between the bottom plane and the sphere's surface is below $D_p$, we let the sphere settle freely, no longer prescribing the velocities.
Assuming that this settling velocity is the final one, the acting hydrodynamic forces would naturally be compensated by gravitational and buoyancy forces.
To avoid an otherwise rapid deceleration before the collision, we model these missing forces by setting an external force onto the particle that corresponds to the negative hydrodynamic force, averaged over the last $\Delta x / u_n^{in}$ time steps.

Based on stability and physical considerations, we choose $\textit{St}_n = 100$ for TW1 and $\textit{St}_n = 300$ for TW2.
We use $D_p=30$ to account for the high Stokes numbers and set $u_n^\text{in}= -0.02$ in lattice units.

The velocities required to compute $\Psi_\text{in}$ and $\Psi_\text{out}$ are evaluated at $t_\text{in} = t_c - 0.05 D_p/u_n^{in}$ and
$t_\text{out} = t_c + 0.05 D_p/u_n^{in}$, where $t_c$ is here defined as the instance in time of maximum penetration depth, i.e., minimum gap size.

\subsubsection{Results and discussion}

\begin{figure}[t]
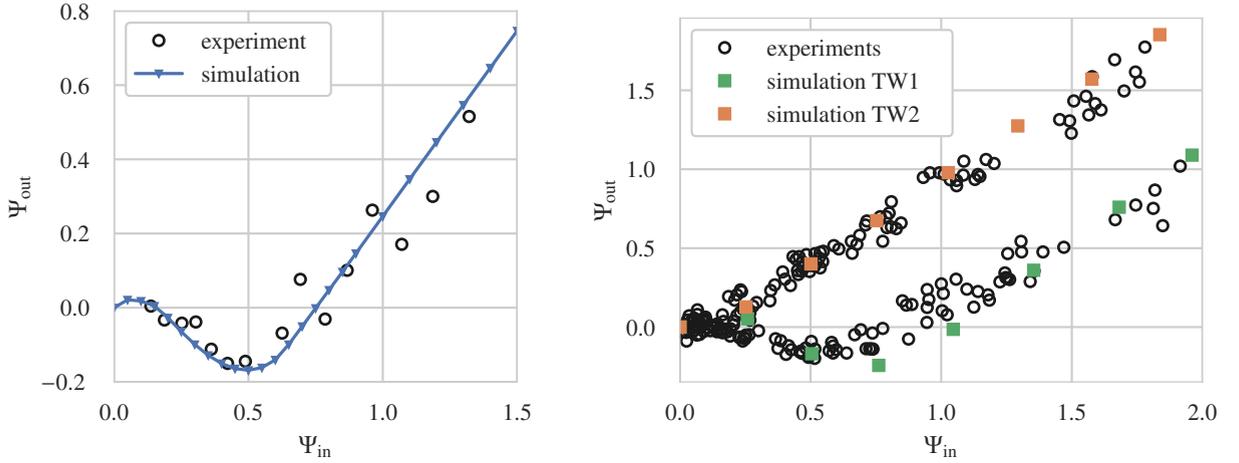

	\centering
	\begin{subfigure}[b]{0.45\textwidth}
		\centering
		\input{figures/oblique_dry_collision.pgf}
		\caption{Dry oblique collision (TD). Experimental data from Ref.~\citenum{foerster1994}.}
		\label{fig:obliqueCollisionDry}
	\end{subfigure}
	\hfill
	\begin{subfigure}[b]{0.53\textwidth}
		\centering
		\input{figures/oblique_wet_collision.pgf}
		\caption{Wet oblique collision (TW1 and TW2). Experimental data from Ref.~\citenum{joseph2004}.}
		\label{fig:obliqueCollisionWet}
	\end{subfigure}
	\caption{Oblique collision of sphere with wall. Physical parameters as in Tab.~\ref{tab:obliqueCollision}.}
	\label{fig:obliqueCollision}
\end{figure}

The dry oblique collision results are shown in Fig.~\ref{fig:obliqueCollisionDry}, where a very good agreement with the experimental data from Ref.~\citenum{foerster1994} can be seen.
This indicates that our tangential collision model is capable of correctly detecting sticking and slipping regimes, which is a crucial aspect for accurate results, as noted in Ref.~\citenum{biegert2017}.

The outcome of the wet oblique collision simulations can be seen in Fig.~\ref{fig:obliqueCollisionWet}, for both cases, TW1 and TW2.
Both cases show a rather distinct collision behavior in the experiments of Ref.~\citenum{joseph2004}, mainly originating from the different friction coefficients.
Again, the simulations are able to reproduce both cases and yield results that match the experimental measurements very well.

We also carried out these simulations without the tangential lubrication correction model from Eqs.~\eqref{eq:lubrication_correction_force_tangential_trans}--\eqref{eq:lubrication_correction_torque_tangential_rot}, and could observe no influence on the collision behavior.
We assume that this is a direct result of the high Stokes numbers applied in this setup, where lubrication forces generally play a minor role in comparison to other hydrodynamic and collision forces. 

As a result of this test case, we have shown that our tangential collision model can accurately capture oblique collision events without introducing further parameters that would need to be calibrated.
For the here considered case, the often-made assumption of negligible tangential lubrication interactions \cite{kempe2012,costa2015,biegert2017} has been verified.
However, such an assumption should be reevaluated on a case by case basis, depending on the characteristic collision dynamics given by the Stokes number.

\section{Conclusion}

\label{sec:conclusion}

In this work, we developed an efficient four-way coupling scheme for the accurate simulation of particulate flows. 
Our approach relies on a geometrically fully resolved representation of the particle shape.
The fluid phase is simulated with a novel variant of the lattice Boltzmann method.
A special parameter permits the explicit control of the bulk viscosity of the fluid.
For moving particles, we demonstrate that a suitably increased bulk viscosity helps to damp nonphysical oscillations. 
An adaptive variant of this idea
leads to an improved lattice Boltzmann scheme where the bulk viscosity is only
raised in a narrow layer around each particle.
The hydrodynamic interaction between  the fluid and the particles is 
established based on the momentum exchange method. 
For validation, we compared the drag and lift forces acting on
stationary spheres in low and high Reynolds number flows to reference data.

Particle-particle collisions were treated by a discrete element method, thus 
following a modular approach that can be easily extended and adapted.
The parameters of the discrete element method were calibrated and validated 
by a meticulous analysis of the scenario when
a sphere impacts on a plane. 
The test cases include both normal and oblique impact for different fluids.
One critical parameter of the method is the duration of the collision event, which
has significant effect on the collision dynamics.
The evaluation revealed that the collision
time should be chosen large enough to allow an adaption of the fluid to the sudden change 
in the particle velocity. 
Furthermore, we could show that the unresolved lubrication interactions 
must be corrected 
to obtain an accurate collision response.
These corrections are essential in the normal direction, confirming the findings from other simulation approaches~\cite{kempe2012,costa2015,biegert2017}.
In the tangential direction, however, the lubrication forces
played only a minor role, so that it is possible to neglect the corresponding corrections.

Summarizing, the effect of all model parameters was investigated in detail and elaborated 
guidelines for the parameter choice were presented.
Rigorous validations have shown that our coupling scheme can reliably predict fluid-particle and particle-particle interactions for low and high Stokes number cases.
This renders our new method generally applicable for a wide variety of flow simulation scenarios.
We point out that all algorithmic components are designed such that they 
are well-suited for massively parallel execution on supercomputers.
This is achieved by avoiding non-local operations and global synchronizations~\cite{rettinger2017Riverbed,bauer2020}.
Future work will present efficiency results for simulations 
involving a large number of interacting particles.

All relevant details about the simulation setup and reference solutions for all tests have been presented and discussed.
Consequently, this article presents also a pipeline of systematic tests for the calibration and validation 
as it can be applied to all fluid-particle coupling methods. 
This validation procedure can be used to determine appropriate parameter sets and to assess the
accuracy of simulation and coupling approaches.
We explicitly note that the suggested validation experiments are neither limited to lattice Boltzmann methods nor to discrete element methods.
The increasing complexity of the tests helps to expose possible deficits.
With the tests, any deficiency of a scheme can be discovered in the early stages of method development or when a method is implemented in a step-by-step process. 
Here, simulation software is developed driven by systematic tests, thus extending the general principle of test-driven software development to the field of scientific and engineering software.

\begin{appendix}

\section{Moments of the LBM multi-relaxation time model}
\label{app:MRT}

The moment transformation matrix appearing in Eq.~\eqref{eq:MRT} for the here applied multi-relaxation time model~\cite{duenweg2007} is given as
\begin{equation} 
\mathbf{M} = 
\left[\begin{array}{ccccccccccccccccccc}1 & 1 & 1 & 1 & 1 & 1 & 1 & 1 & 1 & 1 & 1 & 1 & 1 & 1 & 1 & 1 & 1 & 1 & 1\\0 & 0 & 0 & -1 & 1 & 0 & 0 & -1 & 1 & -1 & 1 & 0 & 0 & -1 & 1 & 0 & 0 & -1 & 1\\0 & 1 & -1 & 0 & 0 & 0 & 0 & 1 & 1 & -1 & -1 & 1 & -1 & 0 & 0 & 1 & -1 & 0 & 0\\0 & 0 & 0 & 0 & 0 & 1 & -1 & 0 & 0 & 0 & 0 & 1 & 1 & 1 & 1 & -1 & -1 & -1 & -1\\-1 & 0 & 0 & 0 & 0 & 0 & 0 & 1 & 1 & 1 & 1 & 1 & 1 & 1 & 1 & 1 & 1 & 1 & 1\\1 & -2 & -2 & -2 & -2 & -2 & -2 & 1 & 1 & 1 & 1 & 1 & 1 & 1 & 1 & 1 & 1 & 1 & 1\\0 & 0 & 0 & 2 & -2 & 0 & 0 & -1 & 1 & -1 & 1 & 0 & 0 & -1 & 1 & 0 & 0 & -1 & 1\\0 & -2 & 2 & 0 & 0 & 0 & 0 & 1 & 1 & -1 & -1 & 1 & -1 & 0 & 0 & 1 & -1 & 0 & 0\\0 & 0 & 0 & 0 & 0 & -2 & 2 & 0 & 0 & 0 & 0 & 1 & 1 & 1 & 1 & -1 & -1 & -1 & -1\\0 & -1 & -1 & 2 & 2 & -1 & -1 & 1 & 1 & 1 & 1 & -2 & -2 & 1 & 1 & -2 & -2 & 1 & 1\\0 & 1 & 1 & 0 & 0 & -1 & -1 & 1 & 1 & 1 & 1 & 0 & 0 & -1 & -1 & 0 & 0 & -1 & -1\\0 & 0 & 0 & 0 & 0 & 0 & 0 & -1 & 1 & 1 & -1 & 0 & 0 & 0 & 0 & 0 & 0 & 0 & 0\\0 & 0 & 0 & 0 & 0 & 0 & 0 & 0 & 0 & 0 & 0 & 1 & -1 & 0 & 0 & -1 & 1 & 0 & 0\\0 & 0 & 0 & 0 & 0 & 0 & 0 & 0 & 0 & 0 & 0 & 0 & 0 & -1 & 1 & 0 & 0 & 1 & -1\\0 & 1 & 1 & -2 & -2 & 1 & 1 & 1 & 1 & 1 & 1 & -2 & -2 & 1 & 1 & -2 & -2 & 1 & 1\\0 & -1 & -1 & 0 & 0 & 1 & 1 & 1 & 1 & 1 & 1 & 0 & 0 & -1 & -1 & 0 & 0 & -1 & -1\\0 & 0 & 0 & 0 & 0 & 0 & 0 & -1 & 1 & -1 & 1 & 0 & 0 & 1 & -1 & 0 & 0 & 1 & -1\\0 & 0 & 0 & 0 & 0 & 0 & 0 & -1 & -1 & 1 & 1 & 1 & -1 & 0 & 0 & 1 & -1 & 0 & 0\\0 & 0 & 0 & 0 & 0 & 0 & 0 & 0 & 0 & 0 & 0 & -1 & -1 & 1 & 1 & 1 & 1 & -1 & -1\end{array}\right]
\end{equation}
Note, that the second, third and fourth row of $\mathbf{M}$ uniquely define the ordering of the lattice velocities $\boldsymbol{c}_q$.
The corresponding equilibrium moments are 
\begin{equation} \boldsymbol{m}^{eq} = \left( \rho, \  u_x, \  u_y, \  u_z, \  u_x^2 + u_y^2 + u_z^2, \  0, \  0, \  0, \  0, \  2 u_x^2 - u_y^2 - u_z^2, \  u_y^2 - u_z^2, \  u_x u_y, \  u_y u_z, \  u_x u_z, \  0, \  0, \  0, \  0, \  0\right)^\top.
\end{equation}

\section{Definitions of particle and interaction quantities}
\label{app:particleDefinitions}
\begin{figure}[t]
	\centering
	\begin{tikzpicture}[]
	\coordinate[label=left:$\boldsymbol{x}_{p,i}$] (xpi) at (2,2);
	\coordinate[label=right:$\boldsymbol{x}_{p,j}$] (xpj) at (5.4,2.5);
	\coordinate[label=above:$\boldsymbol{x}_{ij}^{cp}$] (C) at ($ (xpj)!0.5!(xpi) $ );
	\coordinate (nij) at ($0.5*(xpi)-0.5*(C)$ );
	\coordinate (tij) at ([rotate=90]nij);
	
	\draw[darkgray, <->] ($(xpi)+(0,-2)$) -- ++(0,2) node[pos=0.5,left]{$R_{p,i}$};
	\draw[darkgray, <->] ($(xpj)+(0,-2)$) -- ++(0,2) node[pos=0.5,right]{$R_{p,j}$};
	\node[fill=black, circle, inner sep=1.5] at (xpi) {};
	\node[fill=black, circle, inner sep=1.5] at (xpj) {};
	\node[fill=black, circle, inner sep=1.5] at (C) {};
	\draw[orange,thick] (xpi) circle (2);
	\draw[orange,thick] (xpj) circle (2);
	\draw[dashed] (xpi) -- (xpj);
	\draw[-latex, thick] (C) -- ++(nij) node[pos=1,below]{$\boldsymbol{n}_{ij}$};
	\draw[-latex, thick] (C) -- ++(tij) node[pos=1,right]{$\boldsymbol{t}_{ij}$};
	\draw[-latex, thick] (xpi) -- ++(1.8,-1.5) node[pos=1,right]{$\boldsymbol{u}_{p,i}$};
	\draw[-latex, thick] (xpj) -- ++(-0.4,1.3) node[pos=1,left]{$\boldsymbol{u}_{p,j}$};
	
	\coordinate (intersect1) at ($(C)-0.33*(nij)$);
	\coordinate (intersect2) at ($(C)+0.33*(nij)$); 
	\draw[blue] (intersect1) -- ++($-2.7*(tij)$);
	\draw[blue] (intersect2) -- ++($-2.7*(tij)$);
	\draw[thick,<->,blue] ($(intersect1)-2.5*(tij)$) -- ($(intersect2)-2.5*(tij)$) node[pos=0.5,above]{$\delta_{ij,n}$};
	
	\draw[-latex,thick] ($(xpi) + (150:2.3)$) arc (150:95:2.3);
	\node[above] at ($(xpi) + (95:2.3)$) {$\boldsymbol{\omega}_{p,i}$};
	\draw[-latex,thick] ($(xpj) + (-10:2.3)$) arc (-10:35:2.3);
	\node[right] at ($(xpj) + (35:2.3)$) {$\boldsymbol{\omega}_{p,j}$};
	\end{tikzpicture}
	\caption{Schematic representation of two colliding spheres, $i$ and $j$, together with reference quantities required by the collision model.}
	\label{fig:DEM}
\end{figure}
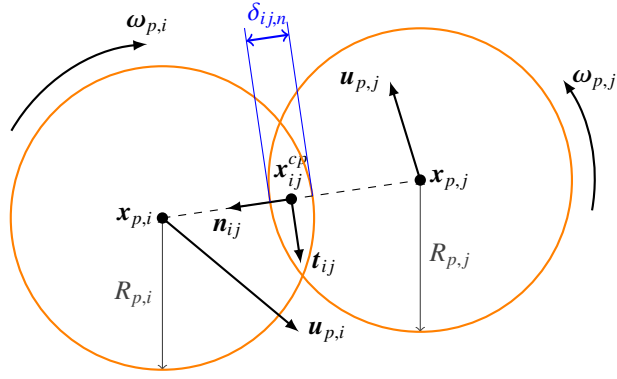

Fig.~\ref{fig:DEM} illustrates two colliding spheres together with their center position $\boldsymbol{x}_p$, translational velocity $\boldsymbol{u}_p$, angular velocity $\boldsymbol{\omega}_p$, and radius $R_p$.
An additional subscript $i$ and $j$ is used to distinguish the quantities of the two spheres.
The velocity of a particle $i$ evaluated at a position $\boldsymbol{x}$ is then
\begin{equation}
\boldsymbol{U}_i(\boldsymbol{x}) = \boldsymbol{u}_{p,i} + (\boldsymbol{x} - \boldsymbol{x}_{p,i}) \times \boldsymbol{\omega}_{p,i}. \label{eq:particleVelocityAtPos}
\end{equation}

During collision, the surfaces of two spheres intersect each other allowing us to define a contact point $\boldsymbol{x}_{ij}^{cp}$ that is located midway between the spheres' centers. 
The unit vector in normal direction from particle $j$ or a wall to particle $i$ in that contact point is
\begin{equation}
\boldsymbol{n}_{ij} = 
\begin{cases}
\frac{\boldsymbol{x}_{p,i} - \boldsymbol{x}_{p,j} }{\|\boldsymbol{x}_{p,i} - \boldsymbol{x}_{p,j} \|},& \quad\text{sphere-sphere,} \\
\boldsymbol{n}_w,& \quad\text{sphere-wall.} \\
\end{cases}
\end{equation}
where $\boldsymbol{n}_w$ is the wall normal, pointing into the open domain. 
The signed surface distance in this normal direction is
\begin{equation}
\delta_{ij,n} = 
\begin{cases}
\|\boldsymbol{x}_{p,i}-\boldsymbol{x}_{p,j}\|-(R_{p,i}+R_{p,j}),& \quad\text{sphere-sphere,} \\
(\boldsymbol{x}_{p,i} - \boldsymbol{x}_w) \cdot \boldsymbol{n}_w -R_{p,i},& \quad\text{sphere-wall.}
\end{cases}
\end{equation}
where $\boldsymbol{x}_w$ is a point at the surface of the plane wall.

Additionally, we define the relative velocity of the particle centers as
\begin{equation}
\boldsymbol{u}_{ij} = \boldsymbol{u}_{p,i} - \boldsymbol{u}_{p,j},
\end{equation}
which can be split into the normal relative velocity
\begin{equation}
\boldsymbol{u}_{ij,n} = (\boldsymbol{u}_{ij} \cdot \boldsymbol{n}_{ij}) \boldsymbol{n}_{ij}
\end{equation}
and the tangential relative velocity
\begin{equation}
\boldsymbol{u}_{ij,t} = \boldsymbol{u}_{ij} - \boldsymbol{u}_{ij,n}.
\end{equation}

With Eq.~\eqref{eq:particleVelocityAtPos}, the relative velocity of the particles' surfaces at the contact point is
\begin{equation}
\boldsymbol{u}_{ij}^{cp} = \boldsymbol{U}_i(\boldsymbol{x}_{ij}^{cp}) - \boldsymbol{U}_j(\boldsymbol{x}_{ij}^{cp}),
\end{equation}
again with the normal and tangential components 
\begin{equation}
\boldsymbol{u}_{ij,n}^{cp} = \boldsymbol{u}_{ij,n}, \quad \boldsymbol{u}_{ij,t}^{cp} = \boldsymbol{u}_{ij}^{cp} - \boldsymbol{u}_{ij,n}^{cp}.
\end{equation}
	
\end{appendix}

\section*{Acknowledgments}

The authors would like to thank Bernhard Vowinckel, Sebastian Eibl, and Ramandeep Jain for valuable discussions.
Furthermore, we gratefully acknowledge the compute resources provided by the Erlangen Regional Computing Center (RRZE).

\bibliographystyle{elsarticle-num}
\bibliography{Library}
	
\end{document}